\documentclass{article}
\usepackage{url}
\usepackage[english]{babel}
\usepackage{algorithm}
\usepackage{algpseudocode}  
\usepackage{amssymb,bm}
\usepackage{graphicx}
\usepackage[letterpaper,top=1.8cm,bottom=1.8cm,left=2cm,right=2cm,marginparwidth=1.75cm]{geometry}
\usepackage{amsmath}
\usepackage{subcaption}
\usepackage{physics}
\usepackage{siunitx}
\usepackage{booktabs}
\usepackage{multirow}
\usepackage{authblk}
\usepackage[numbers]{natbib}
\usepackage{notoccite}
\usepackage[colorlinks=true, allcolors=blue]{hyperref}

\begin{document}

\title{QDsiM: A Noise-Aware Simulation Toolkit for Quantum Diamond Microscope}
\author[1,2]{Satyam Pandey}
\author[1,3]{Abhimanyu Magapu}
\author[1]{Prabhat Anand\thanks{anand.prabhat@tcs.com}}
\author[1]{Ankit Khandelwal}
\author[1]{M. Girish Chandra}

\affil[1]{TCS Research, Tata Consultancy Services Ltd, India}
\affil[2]{Department of Physics, IISER Bhopal, India}
\affil[3]{Birla Institute of Technology and Science, Pilani – Goa, India}

\date{}
\maketitle

\begin{abstract}

The nitrogen-vacancy (NV) center in diamond is a leading solid-state platform for room-temperature quantum magnetometry owing to its long spin coherence times, optical spin initialization and readout, and high sensitivity to magnetic, electric, and thermal perturbations. As NV-based optically detected magnetic resonance (ODMR) systems transition from controlled laboratory environments toward portable and field-deployable sensors, a detailed understanding of realistic noise sources and experimental imperfections becomes essential for optimizing performance and sensitivity.
In this work, we present a comprehensive simulation framework, i.e., a digital twin, for continuous-wave wide-field ODMR in NV-center ensembles. The model is built upon a physically consistent seven-level description of the NV center and incorporates a broad range of experimentally relevant noise and imperfection mechanisms as modular, parameterized components. These include laser and microwave amplitude fluctuations, microwave phase noise, uncertainty in the NV gyromagnetic ratio, spin dephasing, temperature-induced shifts of the ground-state zero-field splitting, surface-induced magnetic field perturbations, and photon shot noise. Power broadening and contrast degradation arising from optical and microwave driving are captured self-consistently through linewidth calculations. Also, the spatial inhomogeneity is modeled via a Gaussian laser intensity profile across the sensing region.
By explicitly linking noise parameters to experimentally accessible quantities such as laser power, microwave power, beam waist, and integration time, the framework reproduces key qualitative and quantitative features observed in experimental ODMR spectra. We validate the simulator through systematic comparisons with reported experimental behaviors and demonstrate its utility by optimizing the contrast-to-linewidth ratio, a central figure of merit for magnetic-field sensitivity. This digital twin provides a versatile tool for experiment design, parameter optimization, and training data-driven methods from ODMR data for magnetic field reconstruction, thereby supporting the development of practical NV-based quantum magnetometers.

\end{abstract}
\section{Introduction}

\begin{figure}[tb!]
    \centering
    \includegraphics[width=0.6\textwidth]{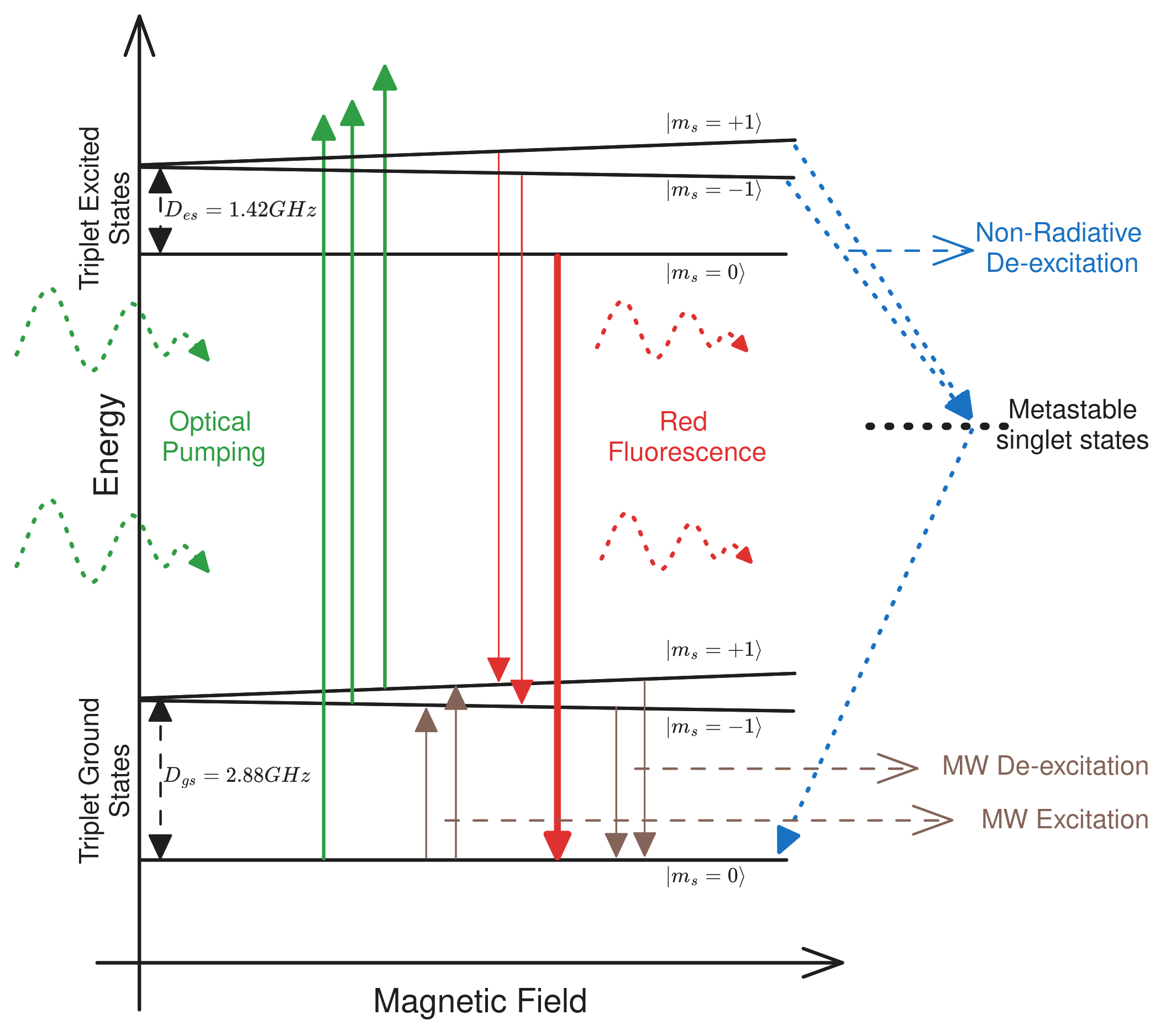}
    \caption{Energy level diagram of the negatively charged NV center in diamond, showing the exact processes that happen when NV is optically pumped along with MW application. It shows the optical and microwave transitions along the ISC through a non-radiative pathway.}
    \label{fig:nvlevels}
\end{figure}
The nitrogen-vacancy (NV) center in diamond has emerged as a cornerstone of solid-state quantum sensing due to its unique combination of optical addressability, long spin coherence times, and operation under ambient conditions~\cite{doherty_nv_review}.
Structurally, the NV center is a point defect formed by a substitutional nitrogen atom located next to a lattice vacancy in the diamond crystal, giving rise to electronic states within diamond’s wide bandgap~\cite{Doherty_2011}.
In its negatively charged state (\( \text{NV}^{-} \)), the defect hosts a spin-triplet ground state whose sublevels can be initialized, manipulated, and read out using a combination of optical and microwave fields. These properties enable precision sensing of magnetic fields, electric fields, temperature, and strain at length scales ranging from nanometers to macroscopic dimensions.
It exhibits remarkable quantum properties, including long spin coherence times, for example, up to approximately 23.6 \textmu s at room temperature in NV ensembles~\cite{10.1063/5.0186997}, and the ability to perform all-optical spin initialization and readout through its spin-dependent photoluminescence (PL) \cite{HARRISON2004245,Tetienne_2012,doi:10.1126/science.276.5321.2012}. 
These optical methods are non-invasive and compatible with room-temperature operation~\cite{PhysRevLett.112.160802}, offering scalability for integration into photonic and quantum-sensing platforms~\cite{chu2015quantumopticsnitrogenvacancycenters}.

The negatively charged NV center (\( \text{NV}^{-} \)) is particularly valuable because of its favorable energy level structure, which lies entirely within the band gap of diamond. 
This structure comprises a spin-triplet ground state and a spin-triplet excited state, enabling optical initialization and readout, as well as coherent microwave control of its spin states. 
The interaction of these spin levels with external perturbations such as magnetic fields~\cite{Taylor_2008}, electric fields~\cite{dolde2011}, temperature~\cite{PhysRevLett.104.070801}, and strain provides a versatile platform for precision sensing in both fundamental research and applied quantum technologies~\cite{CHOE20181066}.

Optically detected magnetic resonance (ODMR) constitutes the primary measurement modality for NV-based magnetometry. The frequencies, linewidths, and contrasts of the dips in the ODMR encode information about the local magnetic field through the Zeeman interaction. For NV ensembles, the presence of four crystallographically distinct orientations allows full vector reconstruction of the magnetic field by analyzing multiple resonance pairs.

Over the past decade, NV-based ODMR has enabled a wide range of applications, including nanoscale magnetometry~\cite{PhysRevB.83.125410}, nuclear magnetic resonance applications~\cite{shi2015}, rotation sensing~\cite{soshenko2020}, electric field sensing~\cite{dolde2011}, quantum registers~\cite{Awschalom2018}, precision time measurement~\cite{hodges2013}, and quantum information applications~\cite{doherty2013}.
While idealized theoretical models capture the essential physics of NV spin dynamics, practical implementations increasingly confront a complex interplay of noise sources and experimental imperfections. Laser and microwave power fluctuations, temperature drift, spin dephasing, strain-induced inhomogeneities, surface-related magnetic noise, and photon shot noise; all degrade ODMR contrast and broaden resonance linewidths, ultimately limiting sensitivity. These effects become particularly significant as NV sensors are miniaturized or deployed outside tightly controlled laboratory environments.

Existing simulation approaches often focus on isolated aspects of NV dynamics, such as idealized spin Hamiltonians, simplified rate-equation models, or phenomenological linewidth descriptions without providing a unified framework that incorporates realistic noise processes in a physically consistent and experimentally parameterized manner. As a result, bridging the gap between ideal theoretical predictions and experimentally observed ODMR spectra remains challenging, especially for ensemble-based wide-field measurements.

In this work, we address this gap by developing a comprehensive digital twin for continuous-wave ODMR in NV-center ensembles. Our approach is based on a seven-level model that explicitly accounts for the ground and excited triplet states, as well as the intermediate singlet manifold, enabling the accurate modeling of optical pumping, intersystem crossing, and spin-dependent fluorescence. Building upon this foundation, we incorporate a wide range of intrinsic and extrinsic noise sources as modular components within the simulation. Importantly, each noise mechanism is parameterized using experimentally accessible quantities, allowing direct correspondence between simulation inputs and laboratory conditions.

The primary objective of this framework is twofold. First, it aims to reproduce experimentally realistic ODMR spectra under a broad range of operating conditions, including the presence of multiple NV orientations, spatial inhomogeneities, and stochastic noise. Second, it provides a quantitative tool for optimizing sensor performance by exploring the trade-offs between contrast, linewidth, and signal-to-noise ratio. By enabling systematic investigation of these effects in silico, the digital twin supports experiment design, parameter tuning, and robust interpretation of ODMR measurements, thereby facilitating the development of reliable, field-deployable NV-based quantum sensors.

The remainder of this paper is organized as follows. Section~\ref{sec:section2} introduces the physical principles of the NV-center, with emphasis on its spin structure and the operational basis of CW ODMR. Section~\ref{simulationframework} presents the overall architecture of the proposed digital-twin framework for NV-center ODMR, outlining its scope and design philosophy. In Section~\ref{sec:sction4}, we develop the theoretical foundations of the model, including the spin-Hamiltonian formalism, the seven-level rate-equation description of optical and spin dynamics, microwave-driven transitions, and linewidth broadening mechanisms. Section~\ref{sec:section5} details the numerical implementation and simulation workflow, describing the algorithmic steps used to compute ODMR spectra for NV ensembles under realistic operating conditions. Section~\ref{sec:section6} systematically examines the role of experimental imperfections and noise sources—such as photon shot noise, laser and microwave power fluctuations, and temperature-induced shifts and analyzes their impact on ODMR contrast and linewidth. Further, the paper discusses the implications of the digital-twin framework for the optimization and deployment of practical NV-based quantum magnetometers while presenting a novel way of denoising ODMR data in Section \ref{sec:discussion}. We conclude by summarizing the key results in Section \ref{sec:conclusion}.

\section{Physical Principles of NV-Center Spin States and ODMR}\label{sec:section2}

\begin{figure}[tb!]
    \centering
    \includegraphics[width=0.8\textwidth]{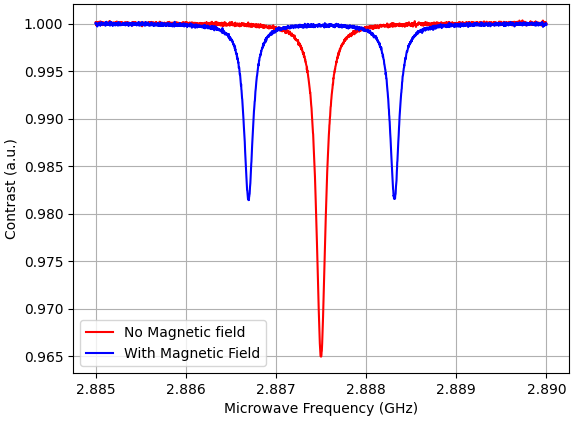}
    \caption{ODMR spectra of a single NV center orientation with and without an external magnetic field.
    The normalized fluorescence intensity is plotted as a function of microwave frequency. 
    The red curve shows the ODMR spectrum in the absence of a magnetic field, exhibiting a single dip centered around \SI{2.87}{GHz}, corresponding to the zero-field splitting between the $m_s = 0$ and $m_s = \pm 1$ ground-state levels. 
    The blue curve shows the spectrum when a static magnetic field is applied along the NV axis. 
    The degeneracy of the $m_s = \pm1$ levels is lifted by the Zeeman effect, resulting in two distinct resonances symmetrically split about the zero-field resonance. 
    This spectral splitting is a key feature used in NV-based magnetometry to infer magnetic field strength and orientation.}
    \label{fig:magnetic}
\end{figure}

The nitrogen-vacancy (NV) center in diamond is a point defect that breaks the perfect tetrahedral symmetry of the diamond lattice and introduces discrete electronic states within its wide bandgap (\SI{5.5}{eV}) \cite{Doherty_2011}. 
The NV center exists in two experimentally observed charge states: neutral (NV$^0$) and negatively charged (NV$^-$)~\cite{doherty2013}. 
A positively charged state (NV$^+$) has also been theoretically predicted, though its experimental confirmation remains challenging~\cite{NVpos}. 
Among these, the NV$^-$ state is of particular interest for quantum sensing and information applications due to its spin-triplet ground state and spin-dependent photoluminescence (PL) properties. \cite{doherty_nv_review}

The axis connecting the substitutional nitrogen atom and the adjacent vacancy defines the symmetry axis of the NV center, which aligns along one of the four crystallographic directions of the diamond lattice. 
This leads to four distinct NV orientations, each experiencing a different projection of any applied external magnetic field. 
The NV$^-$ center contains six electrons—five from the dangling bonds of the nitrogen and neighboring carbon atoms, and one additional electron captured from the lattice~\cite{PhysRevB.53.13441}. 
This configuration forms a spin-1 system with a triplet ground state (\(^3A_2\)), an excited triplet state (\(^3E\)), and intermediate singlet states (\(^1A_1\), \(^1E\)).

As shown in Fig.~\ref{fig:nvlevels}, the ground-state triplet is split into three spin sublevels labeled by the spin projection quantum number \( m_s = 0, \pm1 \). 
In the absence of external fields, the $m_s = \pm 1$ spin sublevels are degenerate and are separated from the $m_s = 0$ state by a zero-field splitting $D_{\mathrm{gs}}$ of approximately $2.87~\mathrm{GHz}$, which originates from spin-spin interactions.
~\cite{10.1063/5.0037414}. 

Under continuous-wave (CW) laser excitation, typically at \SI{532}{nm}, electrons are promoted from the ground to the excited triplet state via spin-conserving transitions. 
Relaxation to the ground state proceeds through two competing channels: a radiative decay that emits red fluorescence, and a non-radiative decay via the intermediate singlet states. 
This non-radiative pathway, known as intersystem crossing (ISC), is spin-selective~\cite{doherty2013} and favors transitions from \( m_s = \pm1 \) to \( m_s = 0 \), enabling both spin polarization and spin-dependent fluorescence. 
These dynamics give rise to two key features: (1) optical spin polarization into the \( m_s = 0 \) state, and (2) optical spin readout based on fluorescence intensity. 
Since the \( m_s = 0 \) state decays primarily through radiative channels, while \( m_s = \pm1 \) states undergo more ISC and emit less light, the spin state can be determined from the PL intensity. 
This spin-dependent photoluminescence forms the basis of NV-based sensing.
\begin{figure}[tb!]
    \centering
    \includegraphics[width=1\textwidth]{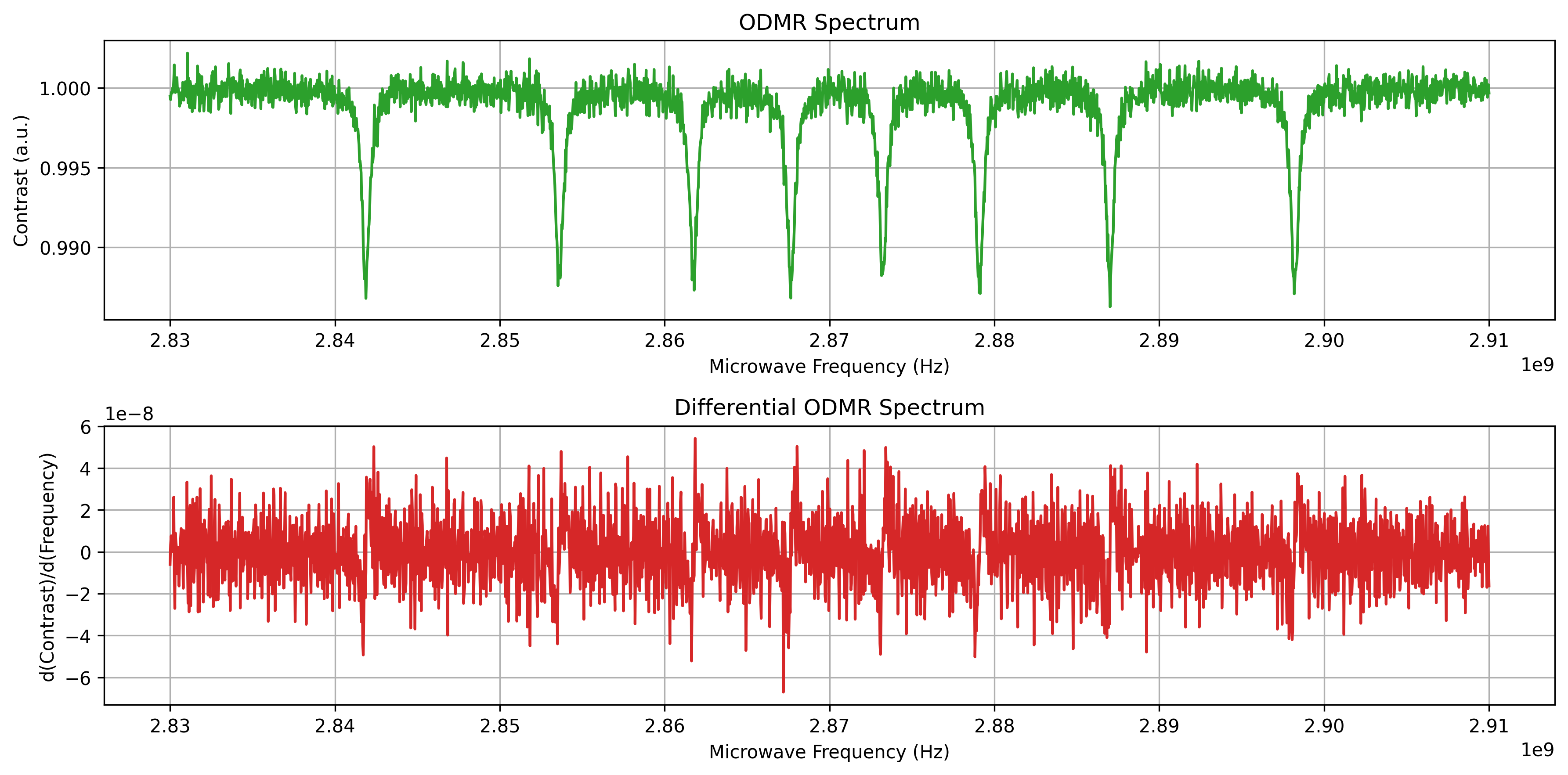}
    \caption{Optically detected magnetic resonance (ODMR) spectrum of an NV ensemble in diamond. 
    Top: The ODMR spectrum displays fluorescence contrast as a function of microwave frequency. 
    Multiple dips are observed due to the Zeeman splitting of the $m_s = \pm 1$ sublevels under an applied magnetic field. 
    NV centers are oriented along four distinct crystallographic axes; the projection of the magnetic field onto each orientation leads to different resonance frequencies, resulting in up to eight distinguishable dips (two per axis, corresponding to transitions $m_s = 0 \rightarrow \pm 1$). 
    Bottom: The differential ODMR spectrum, computed as the numerical derivative of contrast with respect to frequency ($\mathrm{d}C/\mathrm{d}f$), highlights regions of rapid contrast change, improving dip visibility in noisy data.}
    \label{fig:eight_dips}
\end{figure}
In continuous-wave optically detected magnetic resonance (CW ODMR), the NV center is continuously excited by green laser light. At the same time, a microwave field is applied with frequencies swept around \SI{2.87}{GHz}. 
The optical excitation drives the system toward spin polarization into the \( m_s = 0 \) ground state, a process known as optical pumping. 
When the applied microwave frequency is resonant with the energy separation between the $m_s = 0$ and $m_s = \pm 1$ spin states, resonant spin transitions are induced~\cite{doherty2013}. 
These transitions depopulate the \( m_s = 0 \) state and transfer the population to the less fluorescent \( m_s = \pm1 \) states, resulting in a reduction in PL intensity. 
By plotting the normalized PL contrast as a function of microwave frequency, the ODMR spectrum is obtained. 
The characteristic dips in this spectrum correspond to resonance conditions and reveal the spin transition energies, thereby providing a method to analyze the NV energy-level structure through fluorescence intensity measurements.

When a magnetic field is applied with a non-zero component along the NV axis, the degeneracy of the \( m_s = \pm 1 \) states is lifted due to the Zeeman effect \cite{doherty2013}.
This shifts their energy levels in opposite directions, leading to a splitting of the transition frequencies between \( m_s = 0 \) and \( m_s = \pm 1 \). 
As shown in Fig.~\ref{fig:magnetic}, this results in two dips in the ODMR spectrum corresponding to the two transition frequencies. 
The frequency splitting is directly proportional to the component of the magnetic field along the NV axis, providing a precise means of magnetic field detection. 

When considering a diamond sample containing NV centers oriented along all four crystallographic axes, the system is referred to as an NV ensemble. 
For such ensembles, an applied magnetic field will have different projections along each NV orientation, resulting in distinct Zeeman splittings. 
Consequently, the ODMR spectrum can exhibit multiple dips—up to eight when all four orientations experience different field projections, as illustrated in Fig.~\ref{fig:eight_dips}. 
These multiple resonances can be used to reconstruct the full magnetic field vector through a process known as vector magnetometry. The lower panel displays the derivative of the ODMR spectrum, which is employed in practice because the steep zero-crossings of the derivative provide a sharper and more robust indication of the resonance center than the peaks themselves. This method thereby facilitates more precise and reliable identification of the resonance condition \cite{mi16101095}.

In summary, the NV$^-$ center in diamond provides a unique solid-state quantum system with optically addressable spin states. 
Its room-temperature operation and sensitivity to external fields make it a leading platform for quantum sensing and metrology. 
CW ODMR techniques exploit these properties by linking PL intensity to microwave-driven spin transitions, enabling high-resolution measurements of local magnetic fields.
\begin{figure}[tb!]
    \centering
    \includegraphics[width=1\linewidth,]{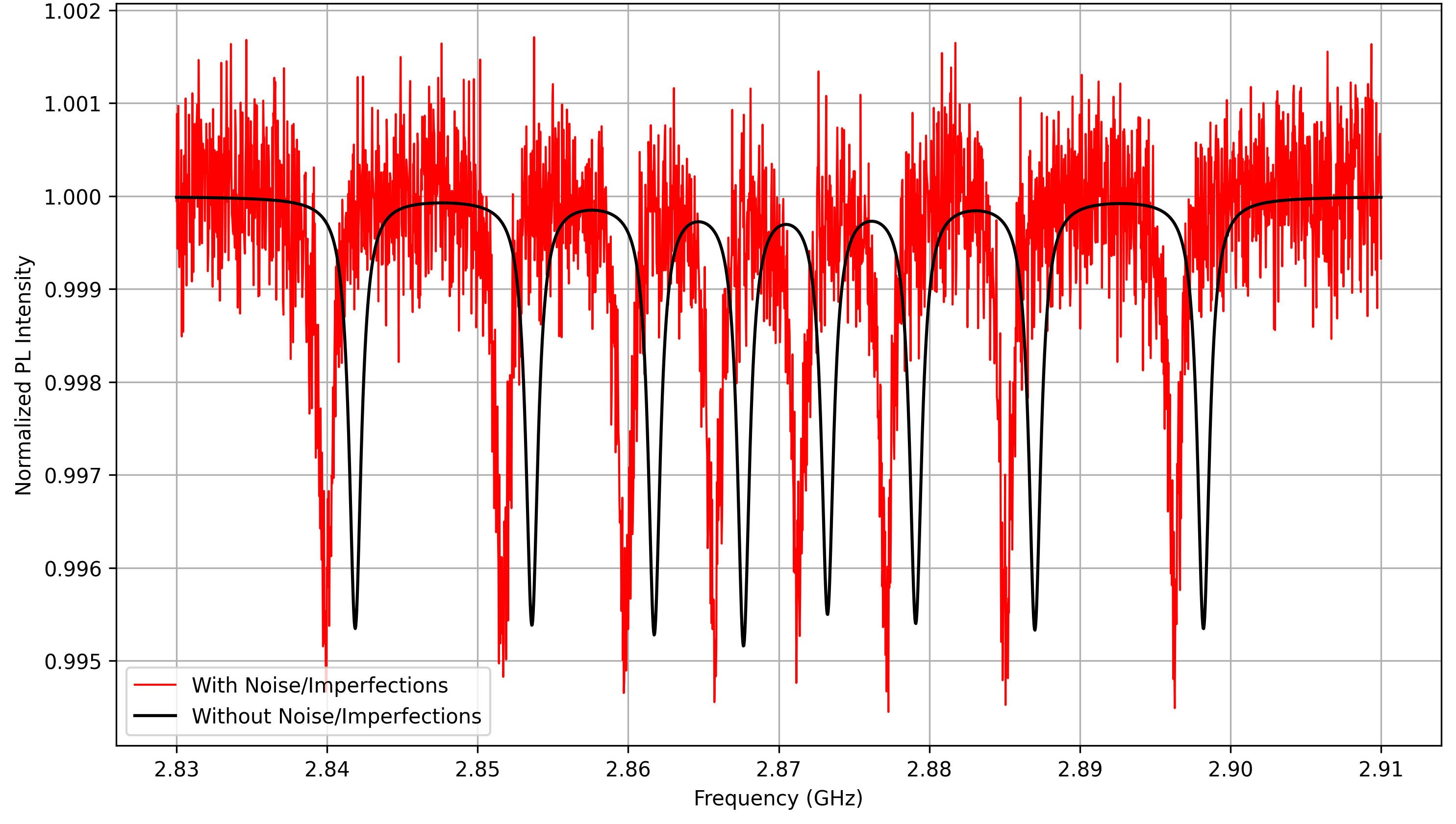}
    \caption{Comparison of optically detected magnetic resonance (ODMR) spectra in the presence and absence of noise and imperfections in an ensemble of nitrogen-vacancy (NV) centers. The black curve represents the ideal ODMR spectrum, free from noise sources, and displays well-resolved resonance dips resulting from the magnetic field-induced splitting of the NV spin sublevels. The red curve illustrates the spectrum with realistic imperfections, including laser and microwave power fluctuations, temperature variations, and uncertainty in the NV gyromagnetic ratio. These effects distort the resonance dips, reduce contrast, and introduce asymmetries, making interpretation more challenging. The comparison highlights the importance of incorporating noise sources into simulations to accurately replicate experimental observations and understand the performance limits of NV-based sensing systems.}
    \label{fig:noise_comparison}
\end{figure}

\section{Architecture for NV-Center ODMR Simulation}\label{simulationframework}

We developed a detailed simulation framework to model the ODMR spectrum of an ensemble of NV centers in diamond. The ensemble includes NV centers aligned along all four possible crystallographic orientations, as typically present in [100]-oriented diamond. Both possible orientations of the nitrogen and vacancy along a given crystallographic axis—namely, NV and VN configurations are incorporated into our analysis. 

A static magnetic field vector is applied uniformly across the ensemble, along with a weak microwave (MW) field that drives transitions between the ground-state spin sublevels. Each NV center is modeled using a seven-level system that includes the ground and excited triplet states, and the intermediate singlet state involved in non-radiative decay. For each microwave frequency point, we solve the optical rate equations to obtain the steady-state populations of the spin levels. These populations are then used to compute the photoluminescence (PL) intensity, both in the presence and absence of the magnetic field, from which the normalized contrast is extracted. This contrast defines the ODMR spectrum.

The resonance dips in the ODMR spectrum are modeled using Lorentzian line shapes, which are commonly employed in continuous-wave (CW) ODMR experiments~\cite{ElElla2017}. Although deviations from a purely Lorentzian profile can occur due to inhomogeneous broadening, this approximation remains valid in most experimental scenarios. For example, El-Ella \textit{et al.}~\cite{ElElla2017} reported that a Voigt profile best fit the experimental data with a 98\% Lorentzian and only 2\% Gaussian contribution, supporting the use of a Lorentzian model for practical purposes, generally limited by power broadening.

The MW-driven transitions are modeled using Fermi’s Golden Rule \cite{galland}, which gives the transition rate under a weak time-dependent perturbation. We adapt this formulation to include a Lorentzian energy spread in the excited state, capturing the linewidth broadening of the contrast dip in the ODMR spectrum. 

To make the simulations more realistic, we incorporate several types of noise, as shown in Fig.~\ref{fig:noise_comparison}. These include photon shot noise, MW amplitude and frequency noise, spin dephasing, and fluctuations in both laser and MW power. Additionally, surface impurity effects are considered. These noise sources are incorporated analytically to produce ODMR spectra that closely resemble experimental observations.

The ODMR spectrum serves as a powerful tool for extracting information about the direction and magnitude of magnetic and electric fields at the position of NV centers. However, its interpretation can be complicated by several factors, including power fluctuations, local intrinsic fields caused by strain, the presence of paramagnetic impurities, surface defects, and variations in NV orientation within the ensemble. These effects introduce complexities that make it challenging to analyze the observed spectral features accurately. Furthermore, identifying optimal sensing configurations is often non-trivial due to the interplay between these factors. Our simulation framework helps analyze these effects and estimate the optimal parameters for experimental design.

The overarching goal of this work is to develop a comprehensive and physically consistent simulation framework capable of accurately reproducing the ODMR response of both single NV centers and NV ensembles under arbitrary external magnetic fields. In the present version of the model, we neglect the contribution of electric fields due to the weak dependence of ODMR splitting and contrast on electric field variations \cite{doherty2013}. 

Ultimately, this framework aims to reproduce ODMR behavior under realistic experimental noise conditions with high fidelity. By doing so, it bridges the gap between ideal theoretical models and practical experimental conditions, providing the foundation for robust, field-deployable NV-based magnetometry and quantum sensing systems.

\section{Theoretical Model and Computational Methods}\label{sec:sction4}
This section systematically outlines the methodology used for simulating the NV-center system and provides the necessary theoretical background for the model employed in this work. It begins with a description of the physical system and the relevant spin Hamiltonian, followed by the calculation of key parameters of interest, such as the linewidth. The interaction Hamiltonian is then introduced, and the model is subsequently extended to describe an ensemble of NV centers. These concepts are refined and expanded in the present study to address the specific objectives of the investigation.

\begin{figure}[tb!]
    \centering

    \begin{subfigure}[b]{0.3\textwidth}
        \centering
        \includegraphics[width=\linewidth, height=0.28\textheight, keepaspectratio]{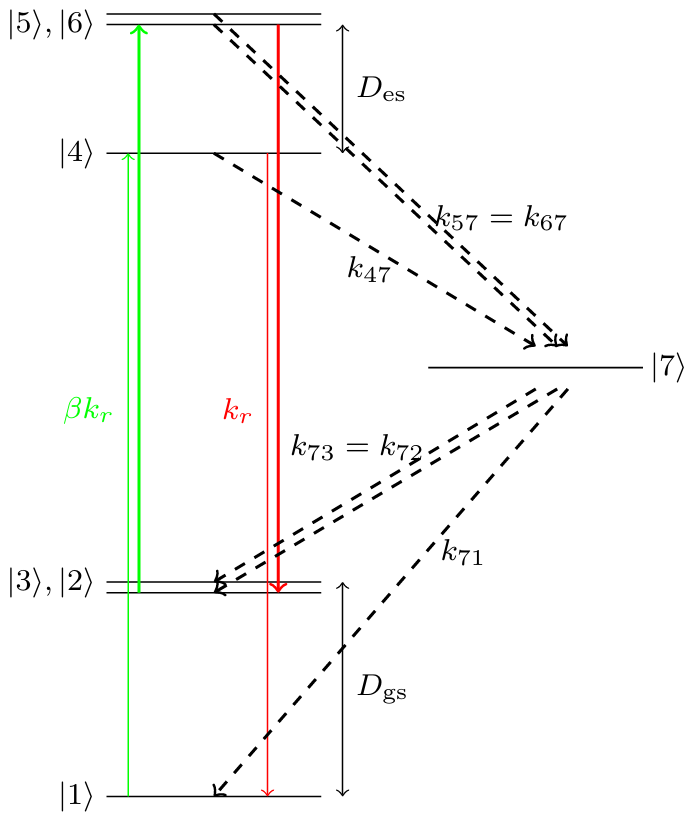}
        \caption{}
    \end{subfigure}
    \hspace{4em}
    \begin{subfigure}[b]{0.3\textwidth}
        \centering
        \includegraphics[width=\linewidth, height=0.28\textheight, keepaspectratio]{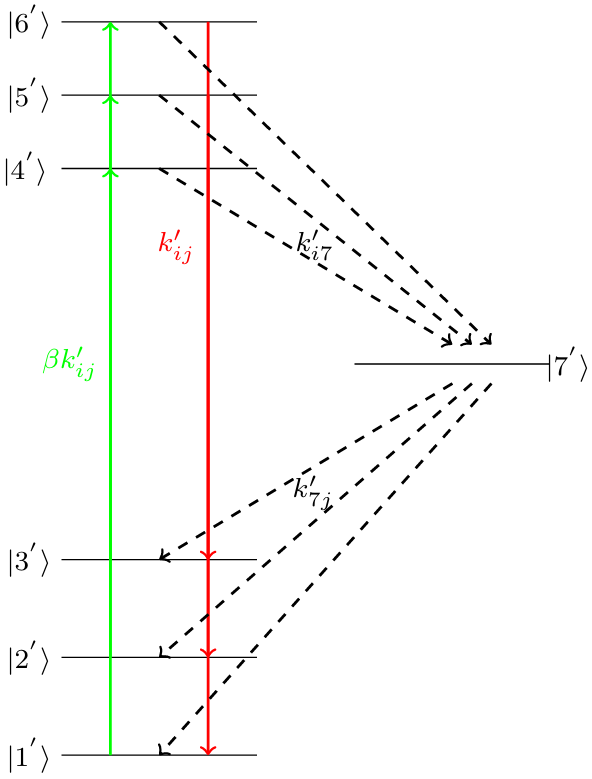}
        \caption{}
    \end{subfigure}
    \caption{(a) Seven-level model of the NV center at zero magnetic field. (b) Magnetic-field-perturbed seven-level system showing Zeeman-induced level mixing and modified transition pathways.}
    \label{fig:sevenlevel}
\end{figure}

\subsection{NV Center as an Effective Spin-1 System}\label{subsec:spin-1}
The negatively charged NV defect is a spin triplet $^3A_2$ in its ground state. As shown in Fig.~\ref{fig:sevenlevel}, the internal spin-spin interaction in the NV decomposes the state into a single spin projection $m_s = 0$ and a doublet degenerate spin projection $m_s = \pm$1 separated by $D_{gs} = 2.87$ GHz at zero magnetic field bias.
When a magnetic field $\boldsymbol{{B}}$ is applied to the defect, the degeneracy is lifted, and $m_s = \pm$1 states are split according to the Zeeman effect \cite{doherty2013}. 
In this scenario, the Hamiltonian $(\mathcal{H}_{gs})$ is written as follows:
\begin{equation}\label{1}
\mathcal{H}_{gs} = h D_{gs} S_z^2 + \mu g \mathbf{{B}} \cdot \mathbf{{S}}
\end{equation}  
In this model, only the ground-state zero-field splitting and the Zeeman interaction with a static magnetic field are taken into account. Furthermore, we adopt the convention that the NV energy levels are shifted such that the $m_s = 0$ state has zero energy, while the $m_s = \pm 1$ states are assigned an energy of $h D_{\mathrm{gs}}$. Here, $\mathbf{{S}}$ are the Pauli spin matrices extended to three dimensions which obey the same algebra as their two dimensional counterpart but operate on a bigger eigenspace, $h$ is Planck's constant, $\mu$ is the Bohr magneton, $g\ (= 2.0028)$ is Landé's g-factor, and $D_{gs} = 2.87 $ GHz is the ground state splitting at zero magnetic field bias.
We have neglected strain-induced splitting \cite{doherty2013}, temperature-induced splitting, and the effect of electric fields - the Stark effect. Additionally, the hyperfine interaction with nearby nuclear spins is neglected, as its contribution is relatively small compared to the dominant electronic spin interactions considered in this model. \cite{doherty2013}

In the case of a single NV center, the transverse field is commonly aligned along the x-axis without loss of generality. However, to facilitate a more comprehensive analysis and enable extension to an ensemble of NV centers oriented along the four distinct crystal axes, we initially consider the Hamiltonian in its most general form. This approach allows us to demonstrate that, even in the most general scenario, the problem can be effectively reduced to a two-parameter model defined by longitudinal and transverse magnetic fields \cite{MITThesis}. Accordingly, the ground-state Hamiltonian is expressed as:

\begin{equation} \label{2}
\mathcal{H}_{gs} = h D_{gs} S_z^2 + \mu_B g B_z S_z + \mu_B g B_x S_x + \mu g B_y S_y = {H}_{gs}^{z} + {H}_{gs}^{\perp}
\end{equation}
where $\mathcal{H}_{gs}^{z} = h D_{gs} S_z^2 + g \mu_B B_z S_z$ and $\mathcal{H}_{gs}^{\perp} = g \mu_B (B_x S_x + B_y S_y)$.

For reference, the spin-1 operators , in the $S_z$ eigenbasis can be represented as:
$$
S_x =
\begin{pmatrix}
0 & \frac{1}{\sqrt{2}} & 0 \\
\frac{1}{\sqrt{2}} & 0 & \frac{1}{\sqrt{2}} \\
0 & \frac{1}{\sqrt{2}} & 0
\end{pmatrix}, \quad
S_y =
\begin{pmatrix}
0 & -\frac{i}{\sqrt{2}} & 0 \\
\frac{i}{\sqrt{2}} & 0 & -\frac{i}{\sqrt{2}} \\
0 & \frac{i}{\sqrt{2}} & 0
\end{pmatrix}, \quad
S_z =
\begin{pmatrix}
1 & 0 & 0 \\
0 & 0 & 0 \\
0 & 0 & -1
\end{pmatrix}.
$$
 The excited state of the NV center is likewise a spin-triplet ${}^3E$ level, exhibiting a structure analogous to that of the ground state, with a zero-field splitting $D_{\mathrm{es}} = 1.42~\mathrm{GHz}$ GHz~\cite{doherty2013}.It shares the same quantization axis and has a gyromagnetic ratio comparable to that of the ground state~\cite{tetienne2012magnetic}. Thus, the excited-state Hamiltonian is given by (\ref{1}) while replacing $D_{gs}$ with $ D_{es}$.
 Additionally, experimental studies have also identified two singlet states (\( ^1E \) and \( ^1A_1 \)) \cite{Rogers_2008, PhysRevB.82.201202}. Theoretical predictions suggest the existence of a third singlet state located between the ground and excited triplet levels \cite{PhysRevB.81.041204}.  Additionally, if we consider two singlet states, the decay rate between the two singlets is swift, and thus they can be treated as a single level \cite{manson2010optically} and hence infrared spectroscopy is not considered in our framework. 
As shown in  Fig.~\ref{fig:sevenlevel} the eigenstates of the ground state are labeled $\ket{1}, \ket{2} \text{and} \ket{3} $ corresponding to the spin projections $m_s$ = 0, -1 and +1. Similarly, the eigenstates of the excited states are labeled $\ket{4}, \ket{5}, \ket{6} $ corresponding to the spin projections $m_s = 0, -1$ and $+1$ and the single metastable level, denoted as \ $|7\rangle$ .

\subsection{Seven-Level Model}
In many practical treatments, the optical excited state is modeled as a separate spin triplet manifold with a zero-field splitting different from that of the ground-state triplet. Although the ground and excited states are coupled optically, their spin–spin and spin–orbit interactions are weak enough that the excited-state manifold can often be considered dynamically decoupled from the ground state, except through these radiative and non-radiative transitions. Consequently, the effective Hamiltonian governing spin dynamics within each manifold is written independently—typically using the ground-state Hamiltonian in (\ref{2}) for the lower triplet, while the excited state enters the seven-level model only through its transition rates and spin-dependent ISC channels that mediate population exchange with the singlet states.
The energy levels of ground, excited, and metastable states are all taken together in a seven-level model of the NV defect \cite{tetienne2012magnetic}.
The seven-level model is a phenomenological, effective model of NV center and the zero-field states (when $B=0$ in Fig. \ref{fig:sevenlevel}(a)) \(\{|i\rangle\}\) forms a basis of eigenvectors of the seven levels. Here, the metastable state is only phenomenological, and it captures the effect of ISC in our simple seven-level model. 
When an external magnetic field is applied, the degeneracy is lifted in the $|2\rangle, |3\rangle$ \text{and} $|5\rangle, |6\rangle$, and we have the states \(\{|i\rangle\}\) are transformed to new seven states \(\{|i'\rangle\}\). Consequently, the new states that arise from the application of an external magnetic field can be expressed as a linear combination of the original states since the original states form a complete basis \cite{tetienne2012magnetic}:
\begin{equation} \label{3}
    \ket{i^{'}} = \sum_{j=1}^{7} \alpha_{ij}(\textbf{B})\ket{j}
\end{equation} 
The state $\ket{i'}$ denotes the $i$th energy level in the presence of an external magnetic field $\mathbf{B}$. This is a perturbed state that includes magnetic-field-induced mixing of the zero-field eigenstates. The expansion coefficient $\alpha_{ij}(\mathbf{B})$ represents the projection of the magnetic-field-perturbed state $\ket{i'}$ onto the $j$th zero-field eigenstate $\ket{j}$, thereby determining the contribution of $\ket{j}$ to $\ket{i'}$. The state $\ket{j}$ itself corresponds to the $j$th eigenstate of the system in the absence of an external magnetic field and forms part of the basis of the seven-level model.




The coefficients $\alpha_{ij}(\textbf{B})$ can be determined using the interaction hamiltonian and are functions of the external static magnetic field applied \cite{lopez2015all}.

Within this model, the zero field transition rates from $\ket{i}$ to $\ket{j}$ are denoted by $k_{ij}$. In this paper, it is considered that the optical transitions are purely spin-conserving and radiative relaxation is spin-independent. Also, the relaxation rate is same for all three $m_s = 0, +1, -1$ states i.e. $k_{41} = k_{52} = k_{63} = k_r$ \cite{tetienne2012magnetic}. The optical pumping rates due to the applied green laser are proportional to the corresponding relaxation rates through $$k_{ji} = \beta k_{ij}$$ for $j = 1, 2, 3$ and $i = 4, 5, 6 $, where $\beta$ is a dimensionless parameter related to the optical pumping rate by \cite{patel2023single}:
\begin{equation} \label{4}
    \beta = \frac{\sigma}{4k_r\times h \times \nu} \times I
\end{equation}
The absorption rate for a \emph{single} NV centre under laser intensity~$I$ is
\begin{equation}
  R_{\text{abs}} \;=\; 
  \frac{\text{absorbed power}}{\text{photon energy}}
  \;=\;
  \frac{\sigma\,I}{h\nu},
  \label{eq:Rabs}
\end{equation}
where
Here, $\sigma$ denotes the absorption cross-section, $I$ represents the laser intensity, and $h\nu$ is the photon energy.

Equation~\ref{eq:Rabs} gives the number of photons absorbed per second (s$^{-1}$) which directly corresponds to the rate of optical excitation ground to excited state:
\[
k_{ji} \;=\; R_{\text{abs}}
           \;=\; \frac{\sigma\,I}{h\nu}.
\]
\[
\beta \;=\; \frac{\sigma\,I}{k_r\,h\nu}.
\]
Hence,
\begin{equation}\label{eq:beta}
    \beta \;=\;
    \frac{\sigma}{4k_r\,h\nu}\,I
\end{equation}
where the factor $4$ accounts for the fact that optical excitation is only possible in one of the four possible NV directions.
$\beta$ is a dimensionless measure of optical pumping strength, comparing the laser-induced excitation rate with the spontaneous radiative relaxation rate $ k_r$.
The intersystem crossing rates depend on the absolute value of $m_s$, i.e. $k_{57} = k_{67}$ and $k_{72} = k_{73}$. The various transition rates are shown in Fig.~\ref{fig:sevenlevel}.

In a similar manner to the eigenstates, the transition rates will evolve as a combination of the zero-field transition rates as follows \cite{Tetienne_2012}:
\begin{equation} \label{5}
k_{ij}^{'}(\mathbf{B}) = \sum_{p=1}^{7} \sum_{q=1}^{7} \left| \alpha_{ip} \right|^2 \left| \alpha_{jq} \right|^2 k_{pq}
\end{equation} 

The excitation due to the green laser causes the spin state populations to evolve with time. The populations can be calculated with the classical rate equation 
\begin{equation} \label{6}
\frac{dn_i}{dt} = \sum_{j=1}^{7} (k_{ji}^{'}n_j - k_{ij}^{'}n_i)
\end{equation} 

However, within the domain of the experiments concerned, the above equation will be solved under steady-state conditions, i.e., $ \frac{dn_i}{dt} \xrightarrow{} 0$. This corresponds to an experimental setup where the NV centre has been excited long enough for all the transient behaviour to diminish. In other words, the rate of excitation from a particular state is equal to that state's relaxation rate.
Our closed seven-level model uses the normalising condition $\sum_{i=1}^{7}n_i = 1$. Thus, to calculate the steady-state spin state population [$\bar{n_i}$], we solve the normalising condition along with the following set of equations 

\begin{equation} \label{7}
0 = \sum_{j=1}^{7} (k_{ji}^{'}n_j - k_{ij^{'}}n_i)
\end{equation} 

The decay through ISC of the excited state to the ground state via the metastable state produces no PL \cite{doherty2013}.  Thus, the total PL is only through the spin-conserving transitions.
Thus, in the general case, the mean PL rate is given by \cite{tetienne2012magnetic}:
\begin{equation} \label{eq:photo}
\bar{R}(\mathbf{B}) = \eta\sum_{i=4}^{6}\sum_{j=1}^{3}\bar{n_i}k_{ij}
\end{equation} 
where $\eta$ is the collection efficiency of the photon detector \cite{lopez2015all}. The relevant transitions that lead to PL are shown in Fig. \ref{fig:sevenlevel} in red colors, and the dashed line represents non-radiative transitions that do not give PL.

\subsubsection{Computing \texorpdfstring{$\alpha_{ij}$}{alpha\_ij}}
For sufficiently small transverse magnetic fields satisfying $\left(\frac{\mu g B_\perp}{\hbar D_{gs}} \ll 1\right)$, the influence of the transverse field on the spin eigenstates can be treated using time-independent perturbation theory (TIPT). The total Hamiltonian of the system is decomposed as follows \cite{MITThesis}:

\begin{equation} \label{9}
H_{Tot} = H_0 + \delta H
\end{equation} \\
with the unperturbed and perturbing Hamiltonians defined as:

\begin{equation} \label{10}
H_0 = h D_{gs} S_z^2 + g \mu_B B_z S_z \
\end{equation} 
\begin{equation} \label{10}
\delta H = g \mu_B (B_x S_x + B_y S_y).
\end{equation} \\
The presence of a finite $B_\perp$ breaks the degeneracy of the states $m_s = \pm1$ and thus non-degenerate TIPT can be used. The unperturbed energy eigenvalues of the system are the usual $S_Z$ eigenvalues $\ket{1}$, $\ket{2}$ and $\ket{3}$. Applying non-degenerate TIPT, the evolved eigenvalues using the above definitions of $H_0$ and $\delta H$ are:

\begin{equation}
  E_{=1} = \hbar D_{gs} + \mu g B_z + \frac{(\mu g)^2 \lvert B_x + i B_y \rvert^2}{2 \left( \hbar D_{gs} + \mu g B_z \right)} + \ldots  
\end{equation}

\begin{equation}
E_0 = \frac{(\mu g)^2 \lvert B_x - i B_y \rvert^2}{2 \left( \hbar D_{gs} + \mu g B_z \right)} + \frac{(\mu g)^2 \lvert B_x + i B_y \rvert^2}{2 \left( \hbar D_{gs} - \mu g B_z \right)} + \ldots
\end{equation}

\begin{equation}
E_{-1} = \hbar D_{gs} - \mu g B_z + \frac{(\mu g)^2 \lvert B_x - i B_y \rvert^2}{2 \left( \hbar D_{gs} - \mu g B_z \right)} + \ldots
\end{equation}

The unperturbed eigenstates are the eigenstates of $S_z$ ($\ket{+^0}, \ \ket{-^0},\ \&\ \ket{0^0}$), and after applying 
TIPT, the evolved eigenstates are found to be:

\begin{equation}
\ket{+} = \ket{+^0} + \frac{\mu g (B_x + i B_y)}{\sqrt{2} \left( \hbar D_{gs} + \mu g B_z \right)} \ket{0^0} + \ldots
\end{equation}

\begin{equation}
\ket{0} = \ket{0^0} - \frac{\mu g (B_x - i B_y)}{\sqrt{2} \left( \hbar D_{gs} + \mu g B_z \right)} \ket{+^0} - \frac{\mu g (B_x + i B_y)}{\sqrt{2} \left( \hbar D_{gs} - \mu g B_z \right)} \ket{-^0} + \ldots
\end{equation}

\begin{equation}
\ket{-} = \ket{-^0} - \frac{\mu g (B_x - i B_y)}{\sqrt{2} \left( \hbar D_{gs} - \mu g B_z \right)} \ket{0^0} + \ldots
\end{equation}

The key physical observables for this system are the energy eigenvalues $E_i$ and the squared magnitudes of the coefficients $|\alpha_{ij}|^2$. From Eq.~14, 15, 16 the relevant magnetic-field components are $B_\parallel$ and $B_\perp$, and all subsequent expressions are formulated in terms of these two quantities so that the extension to NV ensembles is made explicit~\cite{MITThesis}.
From Eq.~14, 15, 16, the coefficients $\alpha_{ij}$ can then be straightforwardly determined by recalling the state mapping $|0\rangle \rightarrow |1\rangle$, $|-\rangle \rightarrow |2\rangle$, and $|+\rangle \rightarrow |3\rangle$. $\alpha_{ij}$ can be calculated by taking the inner product $\lvert \langle j | i \rangle  \rvert $. All non-zero $\alpha_{ij}$ coefficients are listed here:

\begin{alignat*}{2}
\alpha_{11} &= 1, & \qquad \alpha_{44} &= 1, \\
\alpha_{12} &= \frac{\mu g B_\perp}{\sqrt{2} \left( \hbar D_{gs} - \mu g B_\parallel \right)}, 
& \qquad \alpha_{45} &= \frac{\mu g B_\perp}{\sqrt{2} \left( \hbar D_{es} - \mu g B_\parallel \right)}, \\
\alpha_{13} &= \frac{\mu g B_\perp}{\sqrt{2} \left( \hbar D_{gs} + \mu g B_\parallel \right)}, 
& \qquad \alpha_{46} &= \frac{\mu g B_\perp}{\sqrt{2} \left( \hbar D_{es} + \mu g B_\parallel \right)}, \\
\alpha_{21} &= -\frac{\mu g B_\perp}{\sqrt{2} \left( \hbar D_{gs} - \mu g B_\parallel \right)}, 
& \qquad \alpha_{54} &= -\frac{\mu g B_\perp}{\sqrt{2} \left( \hbar D_{es} - \mu g B_\parallel \right)}, \\
\alpha_{22} &= 1, & \qquad \alpha_{55} &= 1, \\
\alpha_{31} &= -\frac{\mu g B_\perp}{\sqrt{2} \left( \hbar D_{gs} + \mu g B_\parallel \right)}, 
& \qquad \alpha_{64} &= -\frac{\mu g B_\perp}{\sqrt{2} \left( \hbar D_{es} + \mu g B_\parallel \right)}, \\
\alpha_{33} &= 1, & \qquad \alpha_{66} &= 1, \\
\end{alignat*}
\vspace{-2.5em}
\begin{equation}
    \alpha_{77} = 1
\end{equation}

\subsubsection{Details of Zero-field Transition Rates}
This model accounts solely for spin-conserving radiative transitions and ISC transitions via the metastable singlet state \cite{tetienne2012magnetic}. The metastable state is considered to interact only with the ground state $\ket{0}$ and the excited states $\ket{+}$ and $\ket{-}$. Moreover, the transition rates are determined by the absolute value of $m_s$, resulting in identical rates for transitions involving the $\ket{+}$ and $\ket{-}$ states. Table \ref{table:rates} shows the numerical values used while creating the simulation tool. $k_{67} = k_{57},\ k_{73} = k_{72}$ as the $\ket{5}$ , $\ket{6}$ and $\ket{3}$, $\ket{2}$ levels are degenerate.

\begin{table}[b!]
\centering
\caption{Transition rates used in the calculations. The values are taken from \cite{tetienne2012magnetic}, and all other transition rates are set to zero.}
\label{table:rates}
\begin{tabular}{cc}
\toprule
\textbf{Transition} & \textbf{Rate [MHz]} \\
\midrule
$k_{r}$   & 63.2  \\
$k_{47}$  & 10.8  \\
$k_{57}$  & 60.7  \\
$k_{71}$  & 0.8   \\
$k_{72}$  & 0.4   \\
\bottomrule
\end{tabular}
\end{table}

\subsubsection{Transition Rates Induced by Microwave Excitation}

In continuous-wave optically detected magnetic resonance (CW-ODMR), spin transitions in the NV center are driven by sweeping an applied microwave (MW) field over a range of frequencies. The MW field can be expressed as:
$$
\mathbf{B}_{\text{MW}}(t) = \mathbf{B}_{\text{MW}} \cos(2\pi \omega t),
$$
where $\mathbf{B}_{\text{MW}}$ is the amplitude vector of the MW magnetic field, and $\omega$ is its angular frequency. To incorporate the effect of this oscillating field into our model, it is coupled to the NV center's ground-state spin Hamiltonian through the interaction Hamiltonian:
\begin{equation} \label{eq:H_int}
H_{\text{int}} = \mu g \left( B_x^{\text{MW}} S_x + B_y^{\text{MW}} S_y + B_z^{\text{MW}} S_z \right),
\end{equation}
where $\mu$ is the Bohr magneton, $g$ is the Landé $g$-factor, and $S_x$, $S_y$, and $S_z$ are the spin-1 operators corresponding to the three spatial directions.

This interaction enables transitions between the ground-state spin sublevels, specifically $\ket{1} \leftrightarrow \ket{2}$ and $\ket{1} \leftrightarrow \ket{3}$. The frequencies at which these transitions occur are determined by the Zeeman energy splittings induced by the applied static magnetic field. Among the spin operators, only the transverse components ($S_x$ and $S_y$) contribute to coherent Rabi oscillations between spin states. In contrast, the longitudinal component ($S_z$) acts as a static field term that shifts the energy levels along the NV axis.

The transition rates between spin states, denoted $T_{12}$ and $T_{13}$, are obtained using time-dependent perturbation theory (TDPT) and Fermi’s Golden Rule \cite{galland}:
\begin{equation} \label{eq:fermi}
T_{if} = \frac{2\pi}{\hbar} \left| \bra{f} H_{\text{int}} \ket{i} \right|^2 \rho(E),
\end{equation}
where $\ket{i}$ and $\ket{f}$ are the initial and final states, and $\rho(E)$ is the density of final states at energy $E$.

To account for inhomogeneous broadening and relaxation processes, $\rho(E)$ is modeled using a Lorentzian distribution centered at the transition energy:
\begin{equation} \label{eq:rhoE}
\rho(E) = \frac{\delta_E/2}{(E - E_f)^2 + (\delta_E/2)^2},
\end{equation}
where $\delta_E$ is the full width at half maximum (FWHM), and $E_f$ is the energy difference between the final and initial states.

Expressing the rate in terms of frequency using $E = h\nu$, where $h$ is Planck’s constant, yields the frequency-domain Lorentzian:
\begin{equation} \label{eq:rhoNu}
\rho(\nu) = \frac{\Delta \nu / (2h)}{(\nu - \nu_f)^2 + (\Delta \nu/2)^2},
\end{equation}
where $\nu_f = E_f/h$ and $\Delta \nu = \delta_E/h$ is the linewidth in frequency units. The corresponding frequency-dependent transition rate is then:
\begin{equation} \label{eq:Tif}
T_{if}(\nu) = \frac{2\pi}{\hbar} \left| \bra{f} H_{\text{int}} \ket{i} \right|^2 \rho(\nu).
\end{equation}

Equation~\ref{eq:Tif} describes how efficiently the MW field drives transitions between the spin sublevels at a given frequency, thereby governing the observed ODMR contrast as a function of MW detuning.
\subsection{Linewidth}

The linewidth observed in the ODMR contrast spectrum is fundamentally limited from Heisenberg's Uncertainty Principle. At the transition frequency, the finite relaxation time of the excited state implies an uncertainty in energy $\Delta E$, leading to a natural linewidth. However, this intrinsic broadening is not the only contribution. Additional effects such as intrinsic dephasing and relaxation due to the NV environment also contribute to the observed linewidth \cite{dreau2011avoiding}. Lastely, optically induced power broadening from both the MW and laser excitation fields majorly affects the apparent linewidth with Lorentzian profile \cite{dreau2011avoiding}. 

At low MW power, transitions between the NV spin states occur at discrete resonance frequencies, yielding sharp, well-defined resonance dips in the ODMR spectrum. As MW power increases, the population of the spin states begins to saturate due to the increase in Rabi frequency $\Omega_R$. When the rate of MW-induced transitions becomes comparable to the intrinsic relaxation rates, saturation occurs, and transitions are driven over a wider frequency range, resulting in power broadening of the resonance peaks.

Similarly, as the power of the excitation laser increases, the rate of optical excitation and the population of NV centers in the excited state also increase. This shortens the optical cycle time, effectively reducing the lifetime of the spin states. According to the uncertainty relation $\Delta E \Delta t \approx \hbar$, this reduction in lifetime leads to an increase in linewidth.

The optical polarization rate $\Gamma_p$, which quantifies the effect of optical pumping, is related to the optical cycling rate and can be expressed as~\cite{dreau2011avoiding}:
\begin{equation}\label{14}
\Gamma_p = \Gamma_p^\infty \times \frac{s}{1+s},
\end{equation}
where $s = P/P_{\text{sat}}$ is the ratio of the applied laser power to the saturation power $P_{\text{sat}}$~\cite{magaletti2024modelling}, and $\Gamma_p^\infty \approx 5 \times 10^6~\text{Hz}$ is the polarization rate at saturation, determined by the lifetime of the metastable singlet state (approximately 200~ns at room temperature~\cite{dreau2011avoiding}).

The saturation power is given by:
\begin{equation}\label{15}
P_{\text{sat}} = \frac{\pi w_0^2}{2} \times I_{\text{sat}},
\end{equation}
where $w_0$ is the laser beam waist, and $I_{\text{sat}}$ is the saturation intensity defined as~\cite{magaletti2024modelling}:
\begin{equation}\label{16}
I_{\text{sat}} = \frac{W_p^{\text{sat}} \, c \, h}{\sigma \, \lambda}.
\end{equation}
Here, $c$ is the speed of light, $\lambda$ is the laser wavelength, $h$ is Planck’s constant, $\sigma$ is the NV center optical cross section, and $W_p^{\text{sat}} = \beta^{\text{sat}} k_r \approx 1.9 \times 10^7~\text{Hz}$~\cite{magaletti2024modelling}.

The relaxation rate of the spin states due to optical pumping can similarly be written as~\cite{dreau2011avoiding}:
\begin{equation}\label{17a}
\Gamma_c = \Gamma_c^\infty \times \frac{s}{1+s},
\end{equation}
where $\Gamma_c^\infty$ is the relaxation rate at saturation. The MW power is modeled using the Rabi frequency
\[
\Omega_R = \frac{\mu g}{h} B_{\text{MW}},
\]
where $B_{\text{MW}}$ is the magnitude of the MW magnetic field.\\
Following the treatment of Dréau et al.~\cite{dreau2011avoiding}, the linewidth of the continuous-wave ODMR resonance of an NV center can be expressed in a phenomenological form that captures both optical and microwave power broadening effects. The final result for the full width at half maximum (FWHM) of the ODMR resonance:
\begin{equation}\label{eq:linewidth}
\Delta\nu = \frac{\Gamma_c^\infty}{2\pi}
\sqrt{
\left(\frac{s}{1+s}\right)^2
+
\frac{\Omega_R^2}{\Gamma_p^\infty \Gamma_c^\infty}
}.
\end{equation}
Here, $\Delta\nu$ denotes the ODMR linewidth (FWHM).The first term inside the square root describes optical power broadening, which increases with laser intensity. The second term accounts for microwave power broadening. Together, these contributions lead to a Lorentzian ODMR lineshape whose width increases with both optical and microwave drive strengths.
\subsection{From Single NV to NV Ensembles}

In a diamond lattice, there are four possible crystallographic orientations along which NV centers can align. The ground-state Hamiltonian is defined in the NV reference frame, where the $z$-axis coincides with the NV axis. However, externally applied fields are typically expressed in the laboratory frame. 

In our case, we consider the four NV orientations, represented by $i = 1, 2, 3, 4$, corresponding to the following direction vectors in the laboratory frame:
\[
\begin{aligned}
    NV_1 &= \left\{\sqrt{\tfrac{2}{3}}, 0, \sqrt{\tfrac{1}{3}}\right\}, \quad
    NV_2 = \left\{0, -\sqrt{\tfrac{2}{3}}, -\sqrt{\tfrac{1}{3}}\right\}, \\
    NV_3 &= \left\{0, \sqrt{\tfrac{2}{3}}, -\sqrt{\tfrac{1}{3}}\right\}, \quad
    NV_4 = \left\{-\sqrt{\tfrac{2}{3}}, 0, \sqrt{\tfrac{1}{3}}\right\}.
\end{aligned}
\]
To model an NV ensemble, we assume that each NV orientation (and its corresponding anti-orientation, NV $\leftrightarrow$ VN) is equally populated in the diamond sample. This assumption is generally valid, although preferential NV alignment can occur in specially grown diamonds. Such cases can be incorporated in the simulation by weighting the photoluminescence (PL) contributions according to the relative abundance of each orientation.\newline
The following steps are used to extend the single-NV model to the NV ensemble case:
\begin{enumerate}
    \item Transform the magnetic field components (static, external, and microwave) into each NV reference frame. The transformation matrix from the laboratory frame to the NV frame depends on the NV axis direction in the lab frame and is given by:
    \[
    R_{NV_1} = 
    \begin{pmatrix}
    0 & 1 & 0 \\
    -\sqrt{\tfrac{1}{3}} & 0 & \sqrt{\tfrac{2}{3}} \\
    \sqrt{\tfrac{2}{3}} & 0 & \sqrt{\tfrac{1}{3}}
    \end{pmatrix},
    \quad
    R_{NV_2} = 
    \begin{pmatrix}
    1 & 0 & 0 \\
    0 & \sqrt{\tfrac{1}{3}} & -\sqrt{\tfrac{2}{3}} \\
    0 & -\sqrt{\tfrac{2}{3}} & -\sqrt{\tfrac{1}{3}}
    \end{pmatrix}.
    \]

    \[
    R_{NV_3} = 
    \begin{pmatrix}
    1 & 0 & 0 \\
    0 & -\sqrt{\tfrac{1}{3}} & -\sqrt{\tfrac{2}{3}} \\
    0 & \sqrt{\tfrac{2}{3}} & -\sqrt{\tfrac{1}{3}}
    \end{pmatrix},
    \quad
    R_{NV_4} = 
    \begin{pmatrix}
    0 & 1 & 0 \\
    \sqrt{\tfrac{1}{3}} & 0 & \sqrt{\tfrac{2}{3}} \\
    -\sqrt{\tfrac{2}{3}} & 0 & \sqrt{\tfrac{1}{3}}
    \end{pmatrix}.
    \]

    \item To obtain the corresponding VN orientations, simply invert the $z$-component of any vector in the NV frame, while keeping the $x$ and $y$ components unchanged.

    \item The same computational procedure as for a single NV center is applied to each of the eight possible orientations (four NV and four VN). For each, the corresponding photoluminescence is computed as $PL_{NV,i}$ and $PL_{VN,i}$, where $i = 1, 2, 3, 4$.

    \item The total ensemble contrast is then obtained by averaging over all orientations:
    \begin{equation}\label{eq:contrast}
        C_{\text{ensemble}}(\nu) =
        \frac{8 C_0 - \left(\sum_{i=1}^{4} PL_{NV,i} + \sum_{i=1}^{4} PL_{VN,i}\right)}{8 C_0}.
    \end{equation}
\end{enumerate}

\section{Numerical Implementation and Simulation Workflow}\label{sec:section5}
The input parameters to the model include the magnetic field vector $\mathbf{B}$ (denoted as \texttt{Bvec} in the code), the detector collection efficiency $\eta$ (\texttt{eta}), the start and end frequencies of the microwave frequency sweep (\texttt{sfreq} and \texttt{ffreq}), and the number of frequency points $N_f$ (\texttt{nfreq}) for a given experimental state. Specifying this \textit{state} of the experimental apparatus can be achieved through the following parameters. The applied microwave power $P_{\mathrm{MW}}$ and its propagation direction, defined by the polar $\theta_{\mathrm{MW}}$ and azimuthal angles $\phi_{\mathrm{MW}}$, determines the spin state excitations in cw-mode. The laser parameters consist of the power $P_{\mathrm{Laz}}$, absorption cross-section $\sigma$, and beam waist w, resulting into an isotropic optical excitation along each NV axis. Noise in the laser and microwave power is introduced through the standard deviations $P_\text{std}$ and $P^\text{mw}_\text{std}$, from their respective distributions. Spin dephasing is accounted for by the parameter $T_2^{*}$ (\texttt{T\_star}), while uncertainty in the NV $g$-factor is represented by \texttt{g\_std}. Finally, the temperature is denoted by $T$ and the integration time for each MW frequency point by $\Delta t$. The value of the experimental parameters used are given in Table~\ref{table:parameter}.

\begin{table}[tb!]
\centering
\caption{Parameter values used in the simulation. Values are based on \cite{magaletti2024modelling} and \cite{segura2019diamond}}
\label{table:parameter}
\begin{tabular}{ll}
\toprule
\textbf{Parameter} & \textbf{Value} \\
\midrule
$P_{\mathrm{MW}}$      & 5–50 dBm \\
$P_{\mathrm{Laz}}$     & 0–1.2 W \\
$w$                  & $10^{-5}\ m$ \\
$\sigma$               & $9 \times 10^{-21}\,\mathrm{m}^2$ \\
$\eta$                 & 1 \\
\bottomrule
\end{tabular}
\end{table}

\subsection{Coordinate Transformation}

Transformations were applied to magnetic field vector to account for four NV orientations in the diamond lattice and get the effective magnetic field in rotated cartesian coordinate system for each NV axis such that z-axis of this new system aligns with NV axis (Algorithm~\ref{algo:transform})

\[
\mathbf{B}_{\text{NV}} = R_{\text{NV}} \mathbf{B}_{\text{lab}}
\]

\begin{algorithm}[H]
\caption{Transform to All NV Frames\label{algo:transform}}
\begin{algorithmic}[1]
\Function{TransformAllNVFrames}{$B_{\text{lab}}$}
    \For{$i = 1$ to $4$}
        \State $R_i \gets \text{RotationMatrix}(i)$
        \State $\mathbf{B_i} \gets R_i \cdot \mathbf{B_{\text{lab}}}$
    \EndFor
    \State \Return $[\mathbf{B_1, B_2, B_3, B_4}]$
\EndFunction
\end{algorithmic}
\end{algorithm}

\subsection{Eigenvalue and Eigenvector Calculation}
We use (\ref{2}) to calculate the ground- and excited-state eigenvalues and eigenvectors. The triplet Hamiltonian includes contributions from the zero-field splitting (ZFS), the Zeeman interaction, and additional shifts arising from temperature variations and spin dephasing (Algorithm \ref{algo:eigen}).

\begin{equation}  
H = h (D(T) + \sigma_i) S_z^2 + \mu_B g_{\text{NV}} \mathbf{B} \cdot \mathbf{S}
\end{equation}
where D(T) is the modified zero field splitting given by (\ref{eq:temp}) after considering the effects of temperature.  

\begin{algorithm}[H]
\caption{Zeeman + ZFS Hamiltonian with Dephasing\label{algo:eigen}}
\label{alg:ham}
\begin{algorithmic}[1]
\Function{Hamiltonian}{$\mathbf{B},T_2^{*},T$}
  \State $\sigma_{D} \gets \mathcal{N}\bigl(0,\;1/T_2^{*}\bigr)$  \Comment{ ZFS Gaussian fluctuation}
  \If{$T>0$} \State $D(T) \gets D_0 + c_1\,n_1(T)+c_2\,n_2(T)$ \Else \State $D(T)\gets D_0$ \EndIf
  \State $H_{\text{ZFS}} \gets h\,[D(T)+\sigma_D]\,S_z^2 $
  \State $H_{\text{Zeeman}} \gets \mu_B g_{\text{NV}} B_z S_z + \mu_B g_{\text{NV}}\,(B_x S_x + B_y S_y)$
  \State $H \gets H_{\text{ZFS}} + H_{\text{Zeeman}}$
  \State $(E,V) \gets \text{eig}(H)$; \quad $E \gets E - \min(E)$
  \State \Return $(E/h,\;V)$
\EndFunction
\end{algorithmic}
\end{algorithm}

\subsection{Interaction Hamiltonian and Transition Strengths}

The interaction Hamiltonian is calculated using (\ref{eq:H_int}). The frequency-dependent transition strengths, $T_{12}$ and $T_{13}$, are then obtained using (\ref{eq:Tif}). The linewidth of the Lorentzian is calculated using (\ref{eq:linewidth}). This procedure results in arrays of $T_{12}$ and $T_{13}$ as functions of frequency, as outlined in Algorithm~\ref{algo:transition_strength}.

\begin{algorithm}[H]
\caption{Compute Transition Strength\label{algo:transition_strength}}
\begin{algorithmic}[1]
\Function{TransitionStrength}{$\vec{B}$, $\vec{B}_{\text{MW}}$}
    \State $H_{\text{int}} \gets \mu_B g_{\text{NV}} \vec{B}_{\text{MW}} \cdot \vec{S}$
    \State $[E, V] \gets \text{Hamiltonian}(\vec{B})$
    \State $T_{12} \gets |\bra{2} H_{\text{int}} \ket{1}|^2$
    \State $T_{13} \gets |\bra{3} H_{\text{int}} \ket{1}|^2$
    \State \Return $T_{12}, T_{13}$
\EndFunction
\end{algorithmic}
\end{algorithm}

\subsection{Linewidth and Optical Pumping Factor}
The linewidth, denoted by $\delta$, is calculated according to (\ref{eq:linewidth}) using the procedure described in Algorithm~\ref{alg:linewidth_computation}. The optical pumping factor, $\beta$, is simultaneously evaluated using (\ref{eq:beta}), along with the magnetic field associated with the applied microwave field. These quantities are then incorporated into the rate-equation model.

\begin{algorithm}
\caption{Compute ODMR Linewidth, Optical Factor, and MW Field}
\label{alg:linewidth_computation}
\begin{algorithmic}[1]
\Function{Linewidth}{$P, P_\text{std}, P^\text{mw}_\text{std}, \sigma, w, P^\text{mw}$}
  \State $\mu_g \gets \mu_B \cdot g_\text{NV}$
  \If{$P_\text{std} = 0$}
    \State $P_\text{laz} \gets P$
  \Else
    \State $P_\text{laz} \gets \text{Normal}(P, P_\text{std})$
  \EndIf
  \State $P^\text{mw} \gets \text{Normal}(P^\text{mw}, P^\text{mw}_\text{std})$
  \State Compute $I_\text{sat}$ and $P_\text{sat}$ using $\sigma$, $w$
  \State $s \gets P_\text{laz} / P_\text{sat}$
  \State $\beta \gets \text{function of } \sigma, \lambda, P_\text{laz}, w$
  \State $\tau_c \gets \text{function of } s$
  \State $\tau_p \gets \text{function of } s$
  \State $B \gets \text{MW\_B}(P^\text{mw})$
  \State $\omega_r \gets \mu_g \cdot B / h$
  \State $\delta \gets \text{function of } \tau_c, \tau_p, \omega_r, \tau_1$
  \State \Return $\delta, \beta, B$
\EndFunction
\end{algorithmic}
\end{algorithm}

\subsection{Population Dynamics and Photoluminescence}
We build a seven-level rate-equation model to compute the steady-state populations $\vec{n}$ and the photoluminescence rate $R$, as given by (\ref{eq:photo}). The steady-state populations are obtained by solving the coupled rate equations as outlined in Algorithm~\ref{algo:population} and are subsequently used to calculate the photoluminescence arising from all allowed transitions through the corresponding transition rates, as outlined in Algorithm~\ref{algo:photo}. Finally, based on the time-step parameter, photon shot noise is incorporated into the total number of collected photons.

\begin{algorithm}
\caption{Compute Steady-State Population\label{algo:population}}
\begin{algorithmic}[1]
\Function{PopulationEquation}{$B, \beta, T_1, T_2$} 
  \State $M \gets \text{PopulationMatrix}(B, \beta, T_1, T_2)$ forms the population matrix
  \State $b \gets$ population vector with one excited state
  \State Solve $M \cdot x = b$ with $x \geq 0$
  \If{solution succeeds}
    \State \Return $x$
  \Else
    \State \Return \textbf{None}
  \EndIf
\EndFunction

\end{algorithmic}
\end{algorithm}
\begin{algorithm}
\caption{Compute Photoluminescence (PL)\label{algo:photo}}
\begin{algorithmic}[1]
\Function{PL}{$B, \beta, \eta, T_1, T_2$}
  \State $K \gets \text{NewRates}(B, \beta, T_1, T_2)$
  \State $n \gets \text{PopulationEquation}(B, \beta, T_1, T_2)$
  \State $R \gets 0$
  \For{$i$ in excited states}
    \For{$j$ in ground states}
      \State $R \gets R + n_i \cdot K_{ij}$
    \EndFor
  \EndFor
  \State $N \gets R \cdot \eta \cdot \Delta t$
  \State Sample $N' \sim \text{Poisson}(N)$
  \State \Return $N'/\Delta t$
\EndFunction

\end{algorithmic}
\end{algorithm}

\subsection{ODMR Contrast Calculation for an NV ensemble with noise}

The final contrast is computed as the normalized difference in photoluminescence before and after the application of microwaves, considering all four NV directions, as given by Eq.~\ref{eq:contrast}. The complete procedure, which employs all previously defined functions to generate an ODMR spectrum for a given set of parameters, is outlined in Algorithm~\ref{algo:odmr}.

\begin{algorithm}[H]
\caption{CW‑ODMR Simulation for an NV Ensemble with Noise}
\label{algo:odmr}
\begin{algorithmic}[1]
\Require Magnetic field in lab frame $\mathbf{B}_{\text{lab}}$,
        collection efficiency $\eta$, microwave direction $(\theta_{\text{MW}},\phi_{\text{MW}})$,\\
        sweep range $f_{\text{start}}\rightarrow f_{\text{end}}$ with $N_{\text{freq}}$ points,\\
        laser power $P$, ${P}_\text{std}$, microwave power $P_{\text{MW}}$, $P_\text{std}^{\text{MW}}$,\\
        laser spot radius $w$, absorption cross‑section $\sigma$,\\
        dephasing time $T_2^{*}$, gyromagnetic‑ratio error $\Delta g$, temperature $T$,\\
        photon‑count integration time $\Delta t$
\Function{ODMR contrast array}{...}

\State $\displaystyle \gamma \gets \frac{\mu_B\,(g_0+\Delta g)}{h}$  \Comment{NV gyromagnetic ratio with uncertainty}
\State $f_k \gets \text{linspace}(f_{\text{start}},f_{\text{end}},N_{\text{freq}})$ \Comment{Microwave sweep grid}
\State $\mathbf{B}^{(i)}_{\text{NV}} \gets
       \textsc{TransformAllFrames}\bigl(\mathbf{B}_{\text{lab}}\bigr)$ for $i=1..4$

\State $C_0 \gets \textbf{0}_{8}$           \Comment{Baseline PL of 4 NVs and their inverted $-B_z$ partners}
\State $C_{\text{amp}} \gets \textbf{0}_{(8,N_{\text{freq}})}$

\For{$i \gets 1$ \textbf{to} $4$}
  \For{$s \in \{+1,-1\}$}                       \Comment{$s$ flips the $z$‑axis once}
    \State $\mathbf{B} \gets (B_x,B_y,s\,B_z) \equiv s\cdot\mathbf{B}^{(i)}_{\text{NV}}$
    \State $(E,V) \gets \textsc{Hamiltonian}\bigl(\mathbf{B},T_2^{*},T\bigr)$ \Comment{Alg.~\ref{alg:ham}}
    \State $(\delta,\beta,B_{\text{MW}}) \gets
           \textsc{Linewidth}\bigl(P,\sigma_P,\sigma_{\text{MW}},\sigma,w,P_{\text{MW}}\bigr)$ \Comment{Alg.~\ref{alg:linewidth_computation}}
    \State $C_0[i,s] \gets \textsc{PL}\bigl(\mathbf{B},\beta,\eta,0,0,\Delta t\bigr)$
    \For{$k \gets 1$ \textbf{to} $N_{\text{freq}}$}
        \State $(\delta,\beta,B_{\text{MW}}) \gets$ \Call{Linewidth}{\dots} \Comment{re‑draw noise every point}
        \State $\mathbf{B}^{\text{MW}}_{\text{NV}} \gets
               \textsc{TransformAllFramesSpherical}(B_{\text{MW}},\theta_{\text{MW}},\phi_{\text{MW}})[i]$
        \State $(t_1,t_2) \gets
               \textsc{TransitionStrength}\bigl(\mathbf{B},
               \mathbf{B}^{\text{MW}}_{\text{NV}},T_2^{*},T\bigr)$     \Comment{Alg.~\ref{algo:transition_strength}}
        \State $t_1 \gets t_1\times\textsc{Lorentzian}\bigl(f_k,E[1],\delta\bigr)$
        \State $t_2 \gets t_2\times\textsc{Lorentzian}\bigl(f_k,E[2],\delta\bigr)$
        \State $C_{\text{amp}}[i,s,k] \gets \textsc{PL}\bigl(\mathbf{B},\beta,\eta,t_1,t_2,\Delta t\bigr)$
    \EndFor
  \EndFor
\EndFor

\State $C_0^{\text{tot}} \gets \displaystyle\sum_{i,s} C_0[i,s]$
\For{$k \gets 1$ \textbf{to} $N_{\text{freq}}$}
   \State $C_{\text{amp}}^{\text{tot}}[k] \gets \displaystyle\sum_{i,s} C_{\text{amp}}[i,s,k]$
   \State $C[k] \gets \dfrac{C_0^{\text{tot}}-C_{\text{amp}}^{\text{tot}}[k]}
                            {C_0^{\text{tot}}}$
\EndFor
\State \Return $C[1{:}N_{\text{freq}}]$
\EndFunction
\end{algorithmic}
\end{algorithm}

\subsection{Reconstructing the magnetic field vector from an ODMR spectrum}
\label{sec:odmr-method}

For an ensemble of NV centres oriented along the four crystallographic directions, the applied magnetic field gives rise to eight distinct resonance frequencies in the ODMR spectrum, corresponding to the two allowed transitions for each NV orientation. Our goal is to extract the three-dimensional magnetic field vector $\mathbf{B}$ from these measured resonance frequencies. The reconstruction method proceeds as follows and as outline in Algorithm~\ref{algo:reconstruction}.

\paragraph{1. Lorentzian fitting of ODMR dips.}
Each ODMR dip is modeled using a single-peak Lorentzian function:
\[
L(\nu; A, \nu_0, \gamma) = A \cdot \frac{\gamma^2}{(\nu - \nu_0)^2 + \gamma^2},
\]
where \( \nu_0 \) is the center frequency, \( \gamma \) is the linewidth, and \( A \) is the amplitude. This allows us to precisely locate the center of each of the eight dips in the spectrum.

\paragraph{2. Resolving pairing ambiguity using a bias field.}
A key challenge in ODMR-based magnetometry is that the eight observed resonances are unlabeled — we do not know which NV axis each pair corresponds to. To eliminate this ambiguity, we can apply a known bias field \( \mathbf{B}_\text{bias} \) that is much stronger than the unknown measured field \( \mathbf{B}_\text{meas} \) (i.e., \( |\mathbf{B}_\text{bias}| \gg |\mathbf{B}_\text{meas}| \)). This bias field shifts the ODMR peaks such that each pair is well-separated and associated with one of the four NV orientations. Thus, the pairing is fixed across all measurements, and we can consistently assign each pair of peaks to its corresponding NV axis.

\paragraph{3. Extracting field projections from frequency splittings.}
For each NV axis \( i \), the magnetic field component along that axis (denoted \( B_{\parallel,i} \)) determines the Zeeman splitting between the \( m_s = \pm1 \) states. This projection is given by:
\[
B_{\parallel,i} = \frac{\nu_{+,i} - \nu_{-,i}}{2\gamma_{\text{NV}}},
\]
where \( \nu_{+,i} \) and \( \nu_{-,i} \) are the two resonance frequencies for NV axis \( i \), and \( \gamma_{\text{NV}} \) is the gyromagnetic ratio of the NV center. Note that this gives the magnitude of the field along axis \( i \), but not its direction.
\begin{figure}[tb!]
    \centering
    \includegraphics[width=0.8\textwidth]{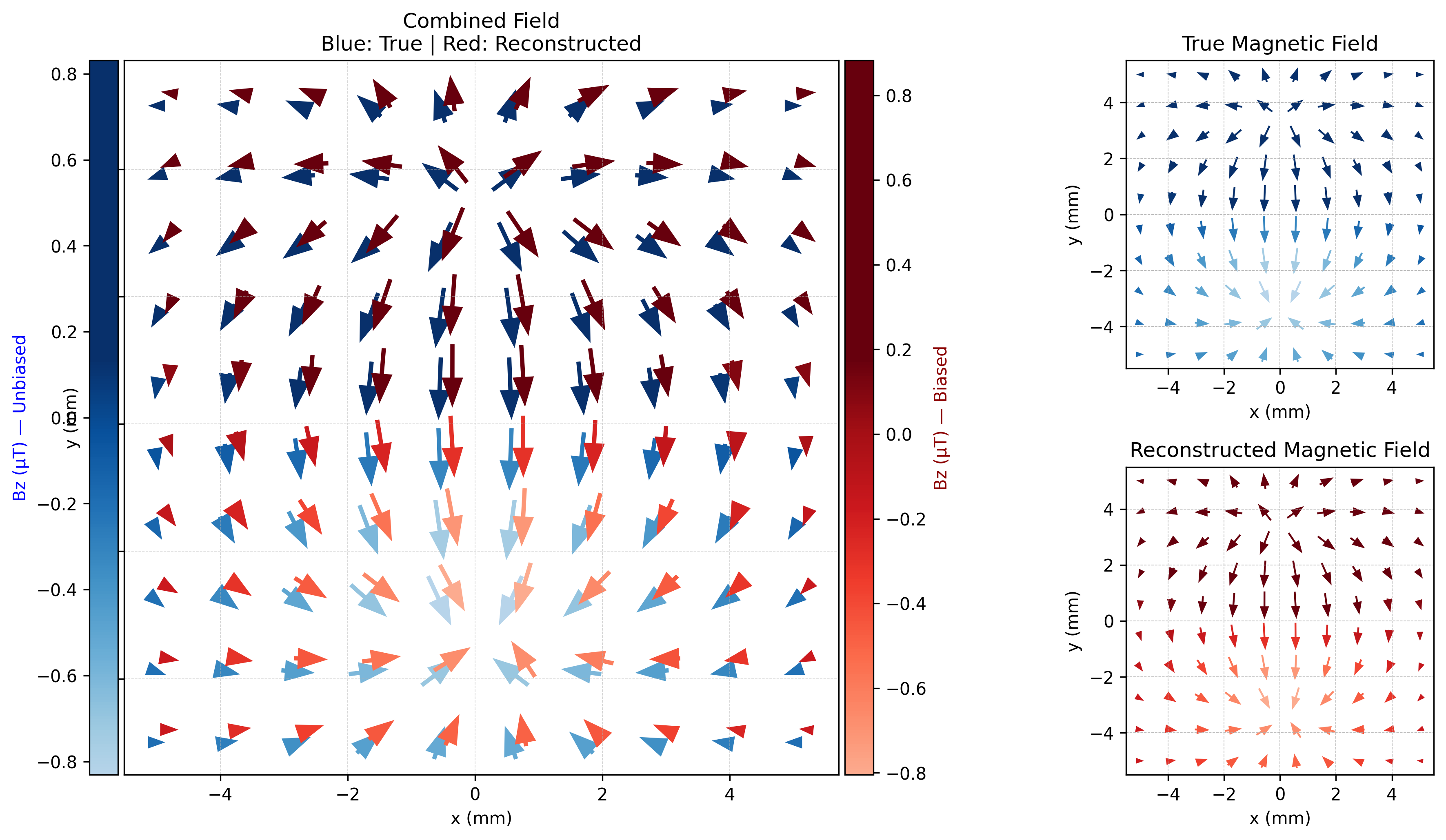}  
    \caption{Comparison of the true magnetic field vectors and the reconstructed field obtained from ODMR data. 
        Left: Overlay of true (blue) and reconstructed (red) in-plane magnetic field vectors. 
        Right-top: True magnetic field. 
        Right-bottom: Reconstructed magnetic field. 
        All vector fields are visualized in the $z=0$ plane. The comparison shows close alignment of the reconstructed vectors with the true field directions, validating the fidelity of the reconstruction method. }
    \label{fig:2drecons}
\end{figure}
\paragraph{4. Determining the sign of each projection.}
Since the ODMR spectrum is symmetric under sign reversal of the magnetic field direction along the NV axis as the splitting only depends on the magnitude of and will be same for parallel and antiparallel components to the NV axis, the frequency difference only tells us \( |B_{\parallel,i}| \). However, in order to reconstruct the full vector \( \mathbf{B} \), we also need to know the direction (parallel or anti-parallel) of the magnetic field component along each NV axis.

To do this, we again use the bias field \( \mathbf{B}_\text{bias} \). Before applying the unknown field \( \mathbf{B}_\text{meas} \), we record the signs of the projections \( \sigma_i = \text{sign}(B^{\text{bias}}_{\parallel,i}) \). Since \( \mathbf{B}_\text{meas} \) is small compared to \( \mathbf{B}_\text{bias} \), the direction of each projection remains unchanged when the two fields are added. Thus, the actual calculated projection of magnetic field is given by:
\begin{align}
 &\nu_{\pm,i} = D_{gs} \pm \gamma_{\text{NV}} |B_{\parallel, i}| \\
 \implies &\sigma_i  \cdot B_{\parallel,i} =  \frac{\nu_{+,i} - \nu_{-,i}}{2\gamma_{\text{NV}}} \notag \\
    \implies & B_{\parallel,i} = \sigma_i \cdot \left(\frac{\nu_{+,i} - \nu_{-,i}}{2\gamma_{\text{NV}}}\right)
\end{align}

\paragraph{5. Reconstructing the magnetic field vector.}
We now have the signed projections of the magnetic field along four known NV axes. Denoting the NV axis directions as unit vectors \( \hat{n}_i \), the four measurements yield:
\[
\hat{n}_1 \cdot \mathbf{B} = B_{\parallel,1}, \quad
\hat{n}_2 \cdot \mathbf{B} = B_{\parallel,2}, \quad
\hat{n}_4 \cdot \mathbf{B} = B_{\parallel,4}, \quad
\hat{n}_3 \cdot \mathbf{B} = B_{\parallel,3}.
\]
These equations form an overdetermined linear system:
\[
\mathbf{N} \mathbf{B} = \mathbf{b},
\]
where \( \mathbf{N} \in \mathbb{R}^{4 \times 3} \) is the matrix of NV axis directions, and \( \mathbf{b} \in \mathbb{R}^4 \) contains the signed projections. Explicitly:
\[
\mathbf{N} =
\begin{bmatrix}
 \sqrt{\frac{2}{3}} & 0 & \sqrt{\frac{1}{3}} \\
 0 & -\sqrt{\frac{2}{3}} & -\sqrt{\frac{1}{3}} \\
 -\sqrt{\frac{2}{3}} & 0 & \sqrt{\frac{1}{3}} \\
 0 & \sqrt{\frac{2}{3}} & -\sqrt{\frac{1}{3}}
\end{bmatrix}, \quad
\mathbf{b} =
\begin{bmatrix}
B_{\parallel,1} \\ B_{\parallel,2} \\ B_{\parallel,4} \\ B_{\parallel,3}
\end{bmatrix}.
\]
We solve this system in the least-squares sense to obtain the full magnetic field vector:
\[
\mathbf{B} = (\mathbf{N}^\top \mathbf{N})^{-1} \mathbf{N}^\top \mathbf{b}.
\]

\begin{figure}[tb!]
    \centering
    \includegraphics[width=0.8\textwidth]{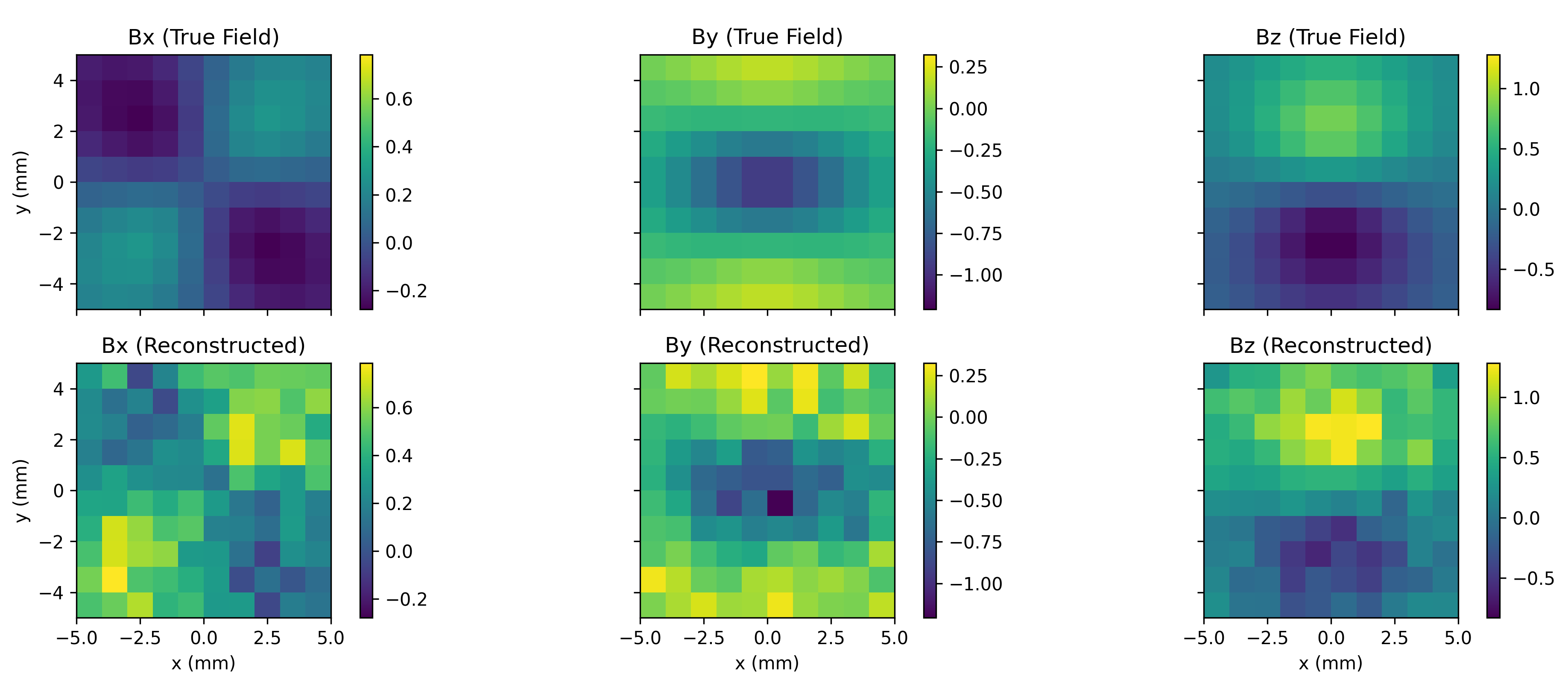}  
    \caption{Heatmaps of the magnetic field components in the $z = 0$ plane. The top row shows the true (unbiased) magnetic field components $\{B_x, B_y, B_z\}$, and the bottom row shows the reconstructed (biased) components obtained from the ODMR-based simulation. Each subplot displays the spatial variation of the respective component in units of \textmu T.}
    \label{fig:heat}
\end{figure}

\paragraph{6. Final output.}
After we get the full magnetic field vector we subtract from it the bias magnetic field to get the desired magnetic field. This procedure yields the full 3D magnetic field vector \( \mathbf{B} \) in the lab frame. The use of a strong bias field ensures that both pairing and direction ambiguities are resolved, allowing accurate vector magnetometry.
The reconstructed magnetic field reflects the influence of errors that shift the resonance peaks, and consequently, it does not exactly coincide with the true field. This deviation is illustrated in the two-dimensional representation shown in Fig.~\ref{fig:2drecons} and in the three-dimensional representation using axis-resolved heat maps in Fig.~\ref{fig:heat}.

\begin{algorithm}[]
\caption{Vector‐Field Reconstruction from an ODMR Spectrum\label{algo:reconstruction}}
\begin{algorithmic}[1]

\Function{Lorentzian}{$x,\,A,\,x_0,\,\gamma$}
    \State \Return $A \cdot \dfrac{\gamma^{2}}{(x-x_0)^{2}+\gamma^{2}}$
\EndFunction
\Statex

\Function{FitSingleLorentzian}{$x,\,y,\,x_0^{\mathrm{guess}}$}
    \State $A^{\mathrm{guess}}\gets\max(y)$
    \State $\gamma^{\mathrm{guess}}\gets (x_{\max}-x_{\min})/200$
    \State $p_0\gets[A^{\mathrm{guess}},\,x_0^{\mathrm{guess}},\,\gamma^{\mathrm{guess}}]$
    \State $p_{\mathrm{opt}}\gets\textproc{curve\_fit}(\textproc{Lorentzian},x,y,p_0)$
    \State \Return $p_{\mathrm{opt}}$
\EndFunction
\Statex

\Function{FindAndFitPeaks}{$\nu,\,C(\nu),\,\texttt{prom},\,\texttt{plotAll}$}
    \State $\Delta(\nu)\gets1-C(\nu)$ \Comment{invert for peak detection}
    \State Detect 8 peaks in $\Delta(\nu)$ with prominence\,=\,$\texttt{prom}$
    \If{number of peaks $\neq 8$}
        \State \Return\;error
    \EndIf
    \For{each peak index $i$}
        \State Estimate half‑width $w_i$ using \textproc{peak\_widths}
        \State $p_{\mathrm{opt}}\gets$ \Call{FitSingleLorentzian}{$\nu,\Delta(\nu),\nu_i$}
        \State Append fitted center $p_{\mathrm{opt}}[2]$ to list
    \EndFor
    \If{\texttt{plotAll}}
        \State Plot $\Delta(\nu)$ and individual Lorentzian fits
    \EndIf
    \State \Return sorted list of 8 centers    
\EndFunction
\Statex

\Function{ReconstructBVector}{$\{\nu_i\},\,\texttt{signs},\,\vec B_{bias},\,\gamma_{\mathrm{NV}}$}
    \State Sort $\{\nu_i\}$ ascending $\to f_1<\dots<f_8$
    \State Pair as $(f_1,f_8),(f_2,f_7),(f_3,f_6),(f_4,f_5)$
    \For{each pair $(f_m,f_p)$}
        \State $|B_{\parallel}|\gets(f_p-f_m)/(2\gamma_{\mathrm{NV}})$
        \State Assign sign from \texttt{signs}; store $B_{\parallel,k}$
    \EndFor
    \State Build vector $\mathbf b=[B_{\parallel,1},B_{\parallel,2},B_{\parallel,4},B_{\parallel,3}]^{\mathsf T}$
    \State Solve least squares $(\mathbf N^{T}\mathbf N)^{-1}\mathbf N^{T}\mathbf b\to\vec B_{\mathrm{meas}}$
    \State $\vec B_{\mathrm{act}} \gets \vec B_{\mathrm{meas}} - \vec B_{bias}$
    \State \Return $\vec B_{\mathrm{act}}$
\EndFunction
\Statex

\Function{GetActualFieldFromODMR}{$\nu,\,C(\nu),\,\vec B_{bias},\,\texttt{signs},\,\texttt{prom},\,\texttt{plot}$}
    \State $\{\nu_i\}\gets$ \Call{FindAndFitPeaks}{$\nu,C(\nu),\texttt{prom},\texttt{plot}$}
    \State $\vec B\gets$ \Call{ReconstructBVector}{$\{\nu_i\},\texttt{signs},\vec B_{bias},\gamma_{\mathrm{NV}}$}
    \State \Return $\vec B$
\EndFunction
\end{algorithmic}
\end{algorithm}

\begin{figure}[tb!]
    \centering
    \includegraphics[width=0.8\textwidth]{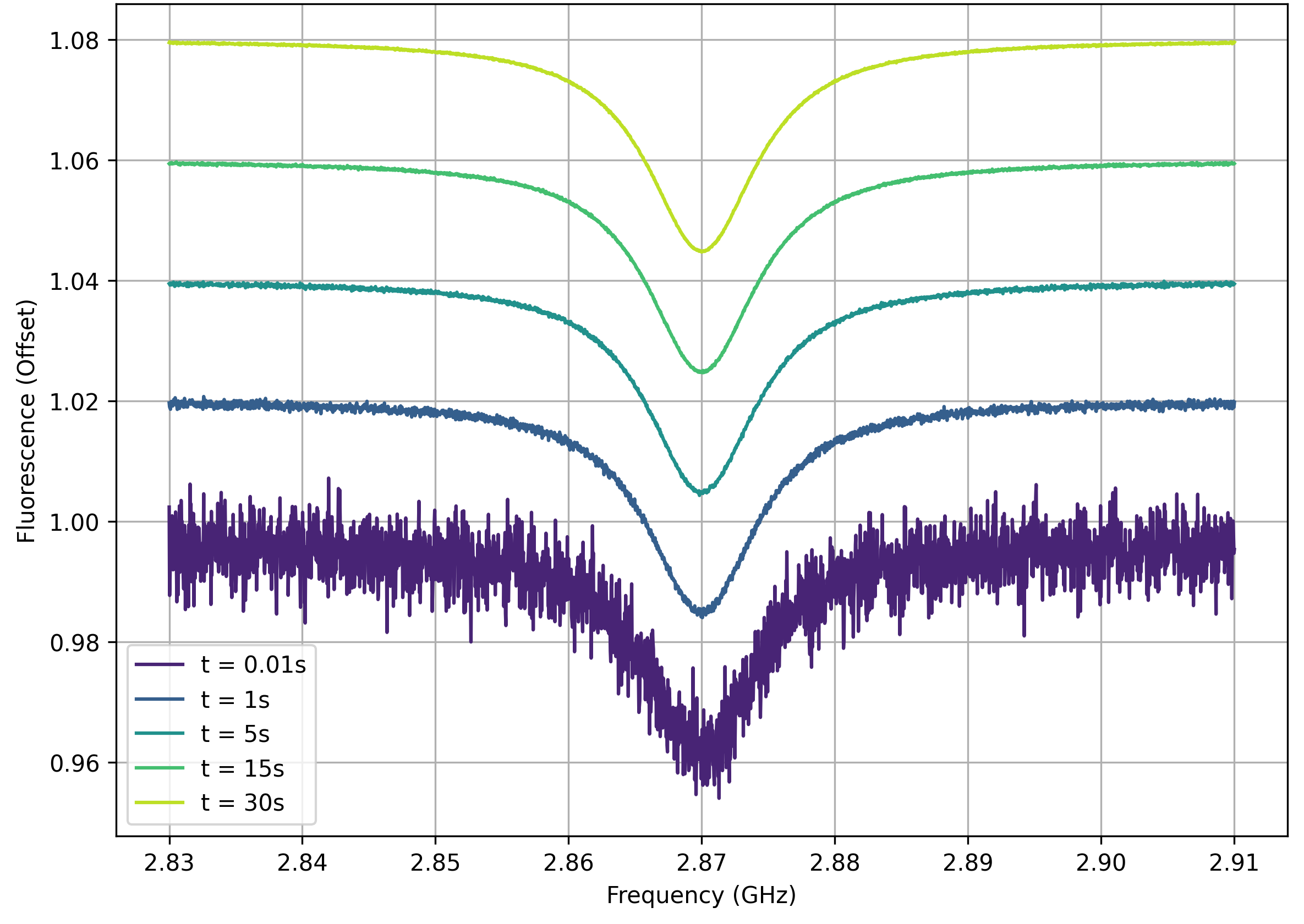} 
    \caption{Plot of ODMR spectra of the NV ensemble offset with each other at different integration times $(t)$.}
    \label{fig:integrationtime}
\end{figure}
\section{Realistic Noise Mechanisms in NV-Center ODMR}\label{sec:section6}


\subsection{Photon Shot noise}

Photon shot noise is a type of random fluctuation that naturally occurs during the detection of light (photons) ~\cite{fox_quantumoptics,mandel_wolf,owen_photonnoise}. 
The detected photon count follows Poisson statistics \cite{goodman_statoptics} with a finite mean value interpreted as light intensity. Accordingly, the photon counts used in the simulation are generated by sampling from a Poisson distribution with a mean equal to the expected photon number~\cite{goodman_statoptics}. 
This procedure introduces photon shot noise–limited sensitivity into our simulation, which depends on the system parameters discussed above.

The fluorescence rate is computed using (\ref{eq:photo}), which gives the photon count per unit time. 
Multiplying this rate by the integration time yields the ideal photon count, \( N_{\text{ideal}} \). 
The integration time ($\Delta t$) represents the duration over which photons are collected at each frequency point of the ODMR spectrum. 
We use \( N_{\text{ideal}} \) as the mean parameter of the Poisson distribution and randomly sample from this distribution to obtain the actual photon count, thereby incorporating photon shot noise into the simulation.

\begin{equation*}
        N_{ideal} = R \Delta t
\end{equation*}
\begin{equation*}
    N_{actual} \sim \mathrm{Poisson}(N_{ideal})
\end{equation*}
Then, the actual photon count is used to calculate contrast. We can see the effect of the various integration times used for the ODMR in Fig.~\ref{fig:integrationtime}.

\subsection{Laser power fluctuations}
An important source of noise in laser beams is power (or intensity) noise \cite{paschotta2008}. This noise originates from both fundamental quantum processes, such as fluctuations associated with the gain medium and resonator losses, and technical sources including pump instability, mechanical vibrations of the cavity mirrors, and thermal fluctuations in the gain medium \cite{Paschotta2005_2}. The lowest achievable limit of laser intensity noise is set by shot noise. 

The experimentally observed power spectral density (PSD) of the laser intensity noise exhibits a distinct peak, as reported in \cite{Paschotta2005}. To reproduce the corresponding statistical fluctuations for the simulation, the peaked PSD was first modeled using a Lorentzian function. This spectrum was then used to generate a simulated time-domain noise signal. Conjugate symmetry was imposed in the frequency domain to ensure that the inverse Fourier transform yield a real-valued signal. A histogram of the resulting amplitude fluctuations was subsequently compared with a standard Gaussian distribution, from which we find that the probability density function (PDF) of the power noise can be well-described by a Gaussian profile.

\begin{figure}[tb!]
    \centering
    \includegraphics[width=0.8\textwidth]{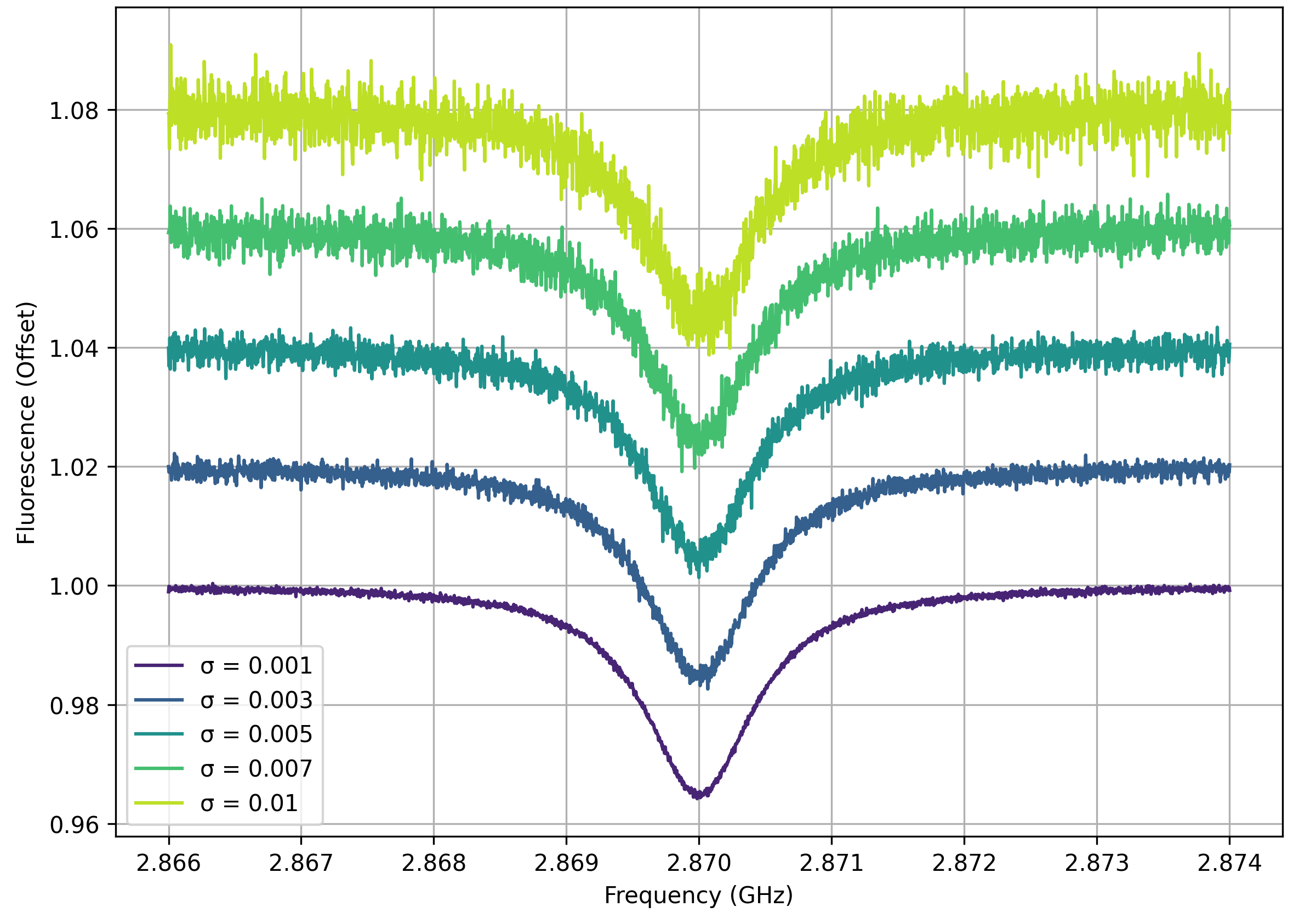} 
    \caption{ODMR spectra of the NV ensemble for varying standard deviations of laser power noise. Spectra are vertically offset with respect to to their respective noise standard deviations.}
    \label{fig:shot}
\end{figure}

In the study of laser intensity fluctuations, it is customary to define the relative intensity noise (RIN)~\cite{Paschotta2007}, which represents the power fluctuations normalized by the mean optical power. The rms value of the power fluctuations can be calculated using the RIN value by integrating it over the laser band and multiplying it by the mean optical power.
\begin{equation}
\mathrm{RIN}(f) \equiv \frac{S_P(f)}{\bar{P}^2}.
\end{equation}
Here, $P(t) = \bar{P} + \delta P(t)$ denotes the instantaneous optical power,
where $\bar{P} = \langle P(t) \rangle$ is the mean optical power and
$\delta P(t)$ represents the stochastic power fluctuations with zero mean.
The quantity $S_P(f)$ is the single-sided power spectral density (PSD) of
$\delta P(t)$. The total root-mean-square (rms) relative intensity fluctuation within a
frequency band $[f_1,f_2]$ is obtained by integrating the RIN:
\begin{equation}
\left(\frac{\delta P}{\bar{P}}\right)_{\mathrm{rms}}
=
\sqrt{
\int_{f_1}^{f_2} \mathrm{RIN}(f)\, df
}.
\end{equation}
It is typically denoted as a percentage of the mean optical power. The usual value is given by 0.1\% to 1\%. So we use a Gaussian distribution with zero mean and 0.5\% of the mean optical power, in our case, the actual applied optical power, as the standard deviation to get the probability distribution function of the noise, which we then add to the laser power to get the actual noisy power value. The effect of the photon shot noise can be observed in Fig.~\ref{fig:shot} where increasing the collection time decreases the noise.

\begin{figure}
    \centering
    \includegraphics[width=0.7\textwidth]{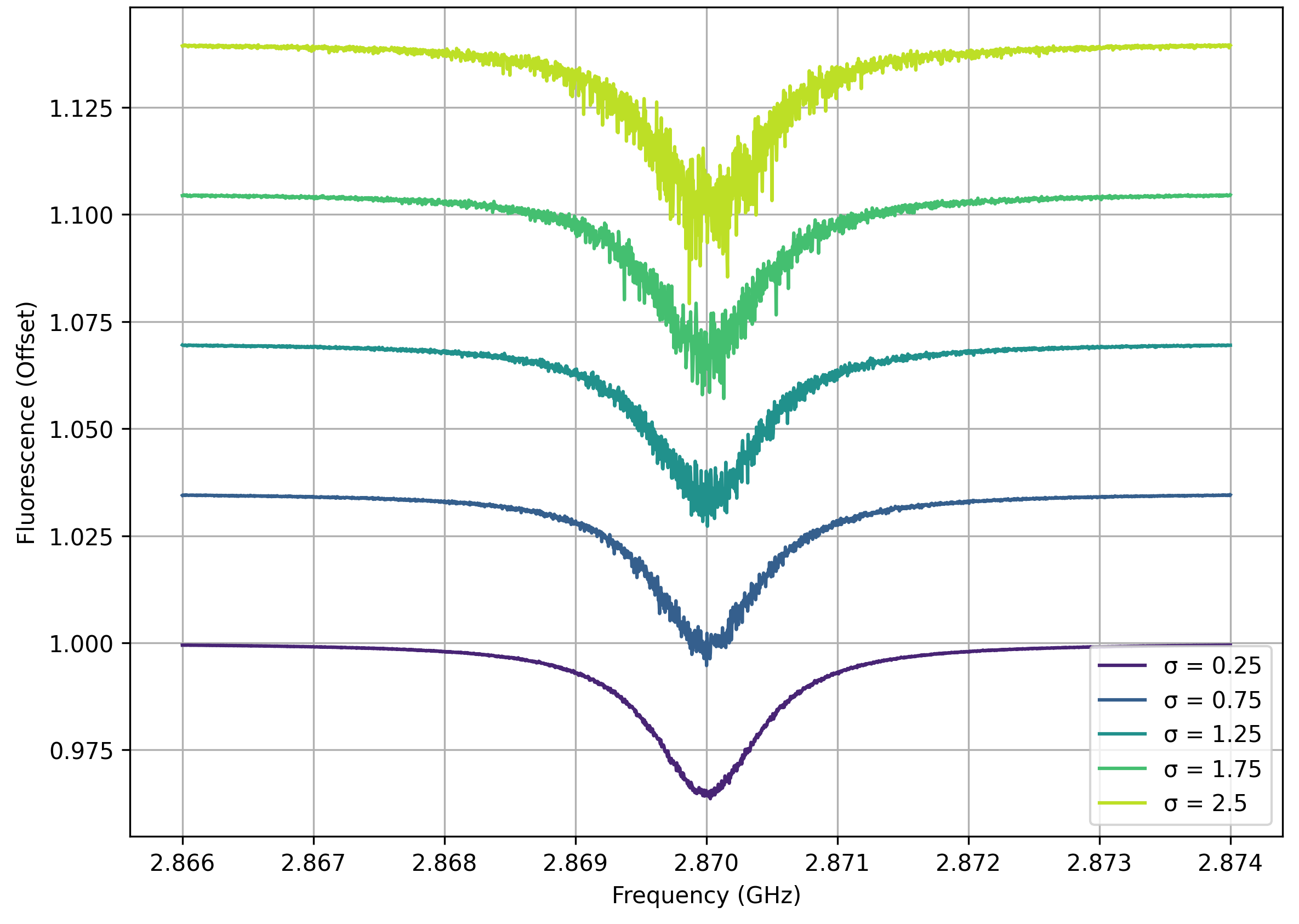} 
    \caption{ODMR spectra of the NV ensemble for varying standard deviations of microwave power noise. Spectra are vertically offset with respect to to their noise standard deviations $(\sigma)$.}
    \label{fig:mw_noise}
\end{figure}

\subsection{Microwave noise}

The applied microwave (MW) field in a CW ODMR experiment is not perfectly stable, but instead exhibits fluctuations in power, phase, amplitude, and frequency arising from a variety of technical noise sources.
The applied microwave power is not perfectly constant over time, but instead exhibits fluctuations and slow drifts. In the absence of a well-established universal statistical model for these fluctuations, we approximate them using a Gaussian distribution centred around the nominal microwave power, with a standard deviation of approximately $0.5\%$ of the specified power. Such fluctuations arise from electronic noise in the microwave source and amplification chain and may include low-frequency components such as flicker (pink) noise. The effect of these power fluctuations on the resulting ODMR spectra is illustrated in Fig.~\ref{fig:mw_noise}.

Fluctuations in the phase of the microwave waveform lead to unintended rotations of the NV spin state. These phase instabilities manifest in the optical readout signal and, if not properly mitigated, are indistinguishable from magnetic-field-induced noise \cite{PhysRevResearch.6.043148}. To account for this effect in our simulation, we introduce an additional noise component by adding a randomly generated value to the magnetic field, constrained within the picotesla (pT) range \cite{PhysRevResearch.6.043148}, corresponding to the magnitude typically affected by such phase noise.

The amplitude of the microwave magnetic field is directly related to the applied microwave power. Consequently, fluctuations in the microwave power lead to variations in the microwave magnetic field amplitude and, therefore, in the Rabi frequency. These fluctuations modify the effective driving strength experienced by the NV spin and result in variations in the ODMR contrast and linewidth. To account for this effect, we use the noisy value of the microwave power when calculating the microwave magnetic field and the associated Rabi frequency.

In an ideal CW ODMR experiment, a microwave signal with a discrete and well-defined frequency is applied at each step. In practice, however, the microwave frequency is subject to inherent fluctuations and exhibits jitter rather than remaining perfectly stable \cite{Barry2020RMP}. To model this effect, we introduce random variations in the microwave frequency at each step of the simulation, thereby capturing the impact of frequency instability on the ODMR response.

\subsection{Temperature fluctuation}
Temperature fluctuations can be caused by various factors, including surrounding equipment, the laser, and even the microwave source's interaction with the NV. These temperature fluctuations result in a shift of the $D_{gs}$, which in turn causes a shift in the center of the ODMR spectrum.
The ground-state zero-field splitting \( D \) of the NV center in diamond is known to depend on temperature due to coupling with lattice phonons. Based on the phonon occupation model, this temperature dependence can be expressed as \cite{PhysRevB.108.L180102}:
\begin{equation}\label{eq:temp}
D(T) = D_0 + c_1 n_1(T) + c_2 n_2(T)
\end{equation}
Here, \( D_0 \) denotes the zero-temperature splitting (\SI{2.87}{GHz}); \( c_1 \) and \( c_2 \) are coupling constants, which are typically negative; and \( n_1 \) and \( n_2 \) represent the mean occupation numbers of the phonon modes with energies \( \Delta_1 \) and \( \Delta_2 \), respectively.

The Bose-Einstein distribution gives the phonon occupation number ($n_i$):

\begin{equation}
n_i(T) = \frac{1}{\exp\left(\frac{\Delta_i}{k_B T}\right) - 1}
\end{equation}

Here, \( k_B \) is the Boltzmann constant; \( \Delta_1 = 58.73 \pm 2.32~\text{meV} \) and \( \Delta_2 = 145.5 \pm 8.4~\text{meV} \); and the coupling constants are \( c_1 = -54.91 \pm 7.35~\text{MHz} \) and \( c_2 = -249.6 \pm 19.4~\text{MHz} \).

\begin{figure}[p]
    \centering
    \includegraphics[width=0.8\textwidth]{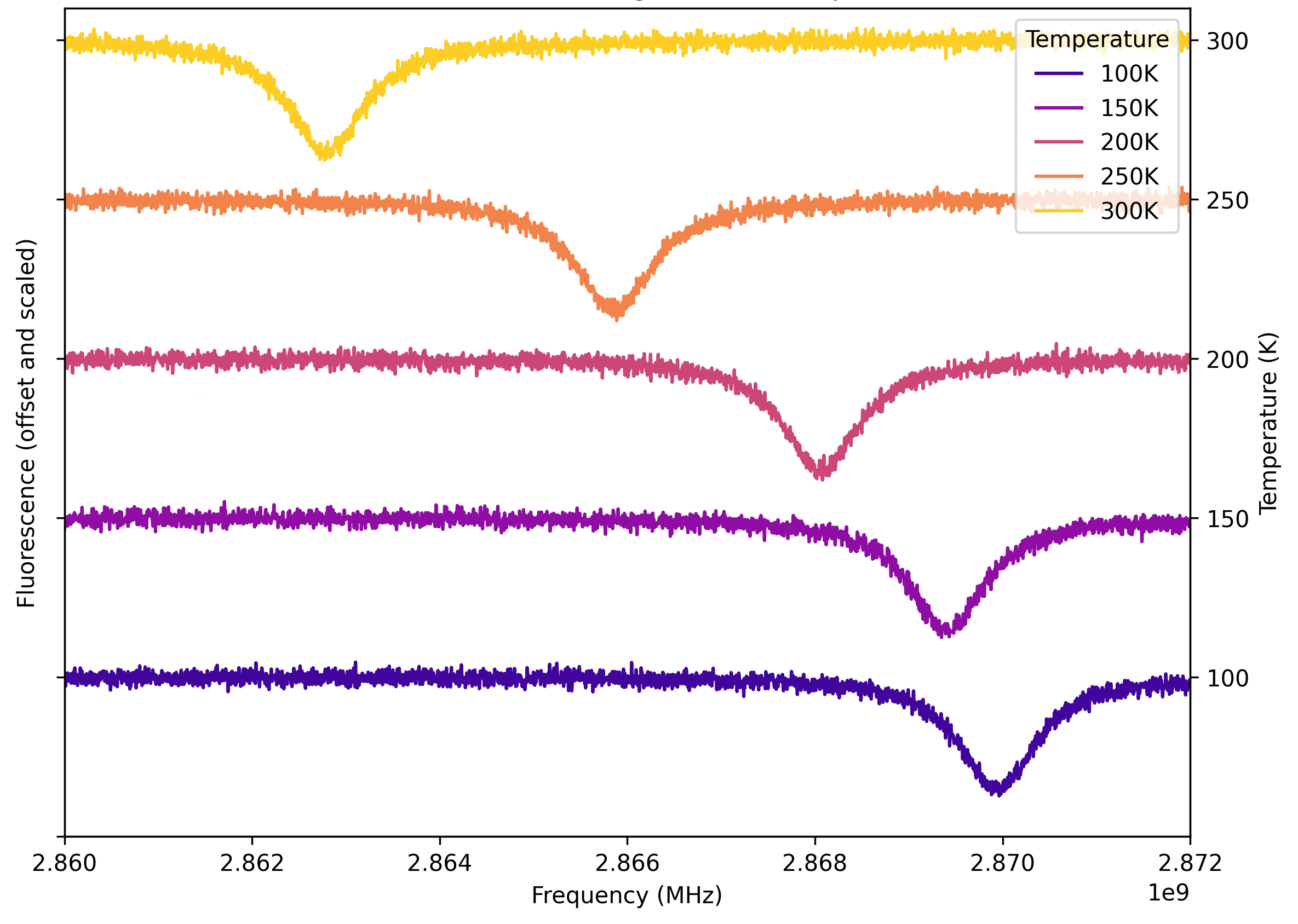} 
    \caption{Plot of the NV ensemble ODMR spectrum at zero magnetic field at different temperatures. The fluorescence values are scaled and normalized for representation purposes.}
    \label{fig:tempshift}
\end{figure}

\begin{figure}[p]
    \centering
    \includegraphics[width=0.8\textwidth]{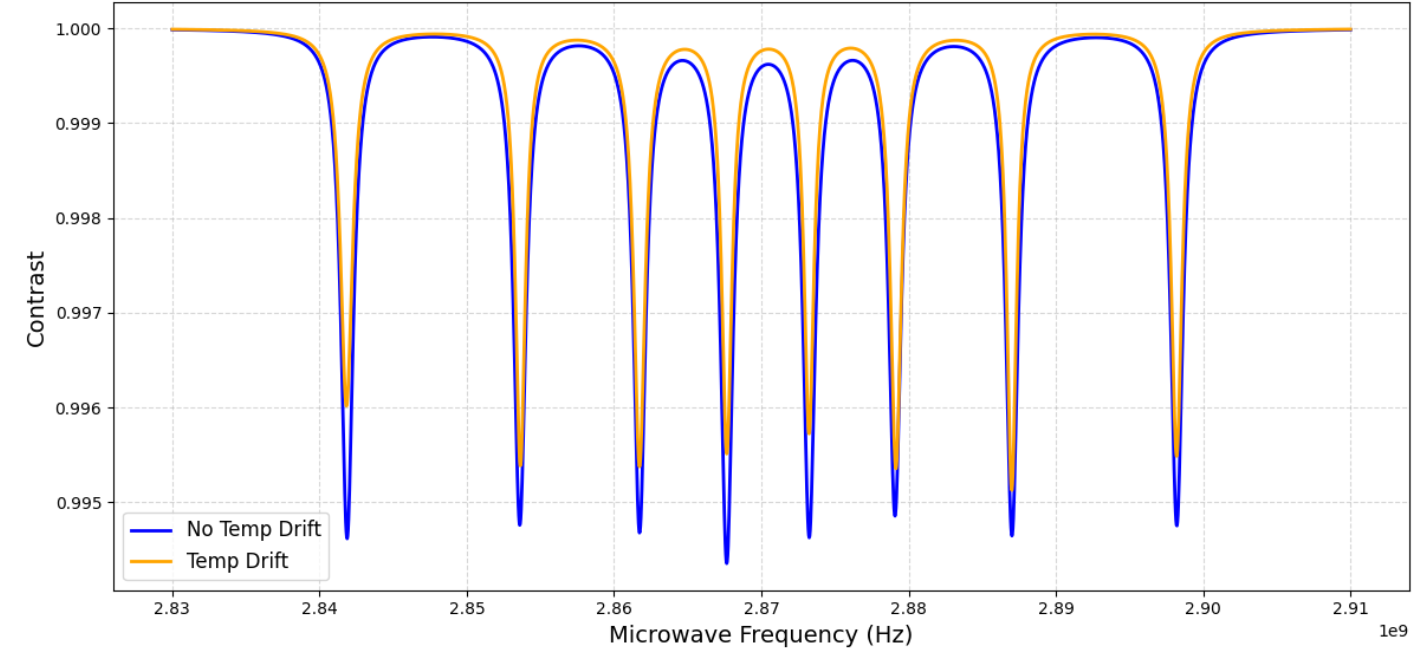} 
    \caption{The plot of the different Lorentzian fitted ODMR with and without temperature drift.}
    \label{fig:drift}
\end{figure}

The shift of the center of the ODMR can be observed in Fig.~\ref{fig:tempshift} The effect of temperature drift can be seen in Fig.~\ref{fig:drift}. The temperature is assumed to drift between $\SI{22}{\celsius}$ to $\SI{28}{\celsius}$ during one sweep of the whole frequency range. This results in a change in the magnetic field vector calculated using the ODMR spectrum thus obtained.
The calculated magnetic field in the presence of temperature drift was found to be as documented in Table~\ref{tab:B_field_comparison}.
\begin{table}[p!]
\centering
\caption{Comparison of reconstructed magnetic field vectors under different conditions.\label{tab:B_field_comparison}}
\setlength{\tabcolsep}{12pt}
\renewcommand{\arraystretch}{1.2}
\begin{tabular}{lccc}
\toprule
\textbf{Condition} & \textbf{$B_x$ (T)} & \textbf{$B_y$ (T)} & \textbf{$B_z$ (T)} \\
\midrule
Temperature Drift   & $6.2302 \times 10^{-6}$ & $3.5449 \times 10^{-6}$ & $1.9461 \times 10^{-6}$ \\
No Drift (Measured) & $5.1241 \times 10^{-6}$ & $3.9328 \times 10^{-6}$ & $2.7906 \times 10^{-6}$ \\
Actual Field        & $5.0000 \times 10^{-6}$ & $4.0000 \times 10^{-6}$ & $3.0000 \times 10^{-6}$ \\
\bottomrule
\end{tabular}
\end{table}

\subsection{Spin dephasing}
The nitrogen-vacancy (NV) center in diamond exhibits a ground-state spin triplet, with the \(|m_s = 0\rangle\) and \(|m_s = \pm1\rangle\) levels split by the zero-field splitting parameter \(D_{gs} \approx 2.87~\text{GHz}\). However, in ensemble measurements or over repeated measurements of a single NV center, inhomogeneities in the local environment (such as strain, electric fields, or a spin bath) lead to inhomogeneous broadening of the resonance, typically described by the spin dephasing time \(T_2^*\)~\cite{article}.

\( T_2^* \) represents the timescale over which coherent superpositions of spin states persist before losing phase coherence. This decoherence arises due to quasi-static local field fluctuations in the environment. It is typically in the range of microseconds. Specifically, in the experiments carried out in \cite{engels}, the dephasing times of the samples ranged from 0.02 to 0.16 $\text{\textmu} s$, with single substitutional nitrogen center densities ranging from 1.8 to 5 ppm.

Such dephasing leads to a distribution of energy levels either across an NV ensemble or over repeated measurements of a single NV center. In the experiments carried out in \cite{PhysRevApplied.21.044051}, these fluctuations follow a Gaussian distribution added to the zero-field splitting parameter \( D_{gs} \). Specifically, we model the fluctuation \(\sigma\) as:

\[
\sigma \sim \mathcal{N}\left(0, \frac{1}{T^*_2}\right)
\]

\begin{figure}[tb!]
    \centering
    \includegraphics[width=0.8\textwidth]{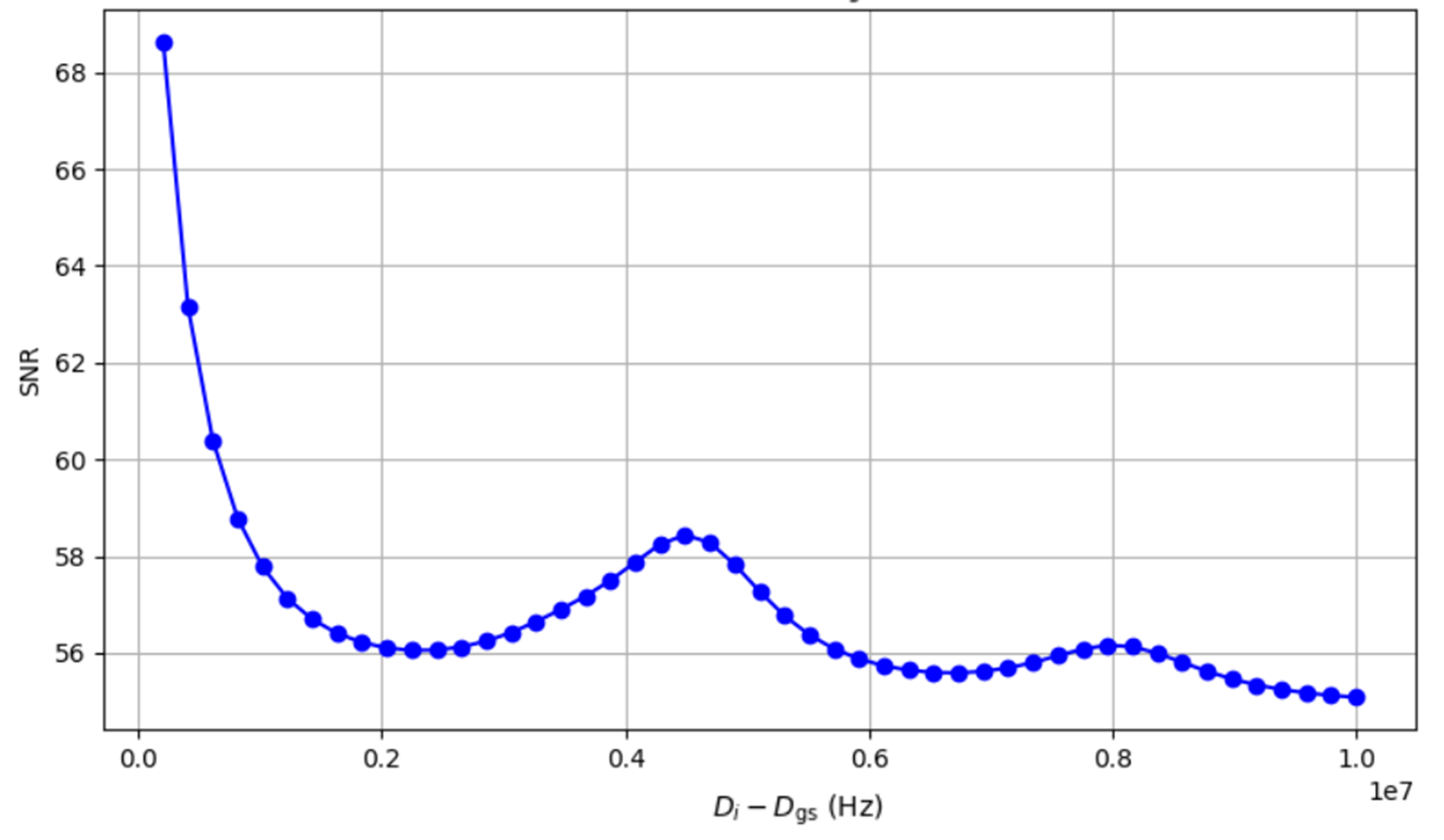}  
    \caption{Plot of signal-to-noise ratio (SNR) of the NV ensemble ODMR spectrm v/s the  value of fluctuations in $D_{gs}$ value due to the spin dephasing . $D_i $ is the total value of zero field splitting after the addition of the fluctuation.}
    \label{fig:D_0}
\end{figure}

This implies the following:
\begin{itemize}
    \item The fluctuation \(\sigma\) has a mean of zero, meaning it is centered around the nominal value of \( D_{gs} \).
    \item The standard deviation of the distribution is \( \frac{1}{T^*_2} \), which corresponds to an effective linewidth of approximately \( 2~\text{MHz} \), as observed experimentally.
\end{itemize}

where \(\frac{1}{T^*_2} \approx 2~\text{MHz}\) characterizes the width of the distribution. This results in the modified NV Hamiltonian:

\begin{equation}
    H_0 = h (D_{gs} + \sigma) S_z^2 + g \mu_B B_z S_z,
\end{equation}
To quantify the change in ODMR signal due to this effect the SNR of the ODMR is calculated for various values of change in $D_{gs}$ due to the spin dephasing as shown in Fig.\ref{fig:D_0}


\subsection{Uncertainty in \texorpdfstring{$T_2^*$}{T2-star}}
The inhomogeneous dephasing time $T_2^*$ of NV centers varies due to several microscopic mechanisms, including local magnetic field fluctuations from $^{13}$C nuclear spins, residual paramagnetic impurities, strain in the diamond lattice, and temperature-dependent effects. These sources differ from one NV center to another and may also fluctuate temporally for a single NV, resulting in uncertainty in the precise value of $T_2^*$.

In bulk diamond, where each NV center experiences a superposition of many weak, independent dephasing contributions, the resulting variation in $T_2^*$ can be modeled statistically. Following the central limit theorem, the sum of numerous independent random perturbations to the local field gives rise to a normally distributed effective dephasing rate. Consequently, we model $T_2^*$ as a random variable sampled from a normal (Gaussian) distribution characterized by a mean $\mu_{T_2^*}$ and a standard deviation $\sigma_{T_2^*}$. This approach captures the natural variability observed experimentally while maintaining a minimal parametrization.

A Gaussian distribution is also consistent with the widely observed Gaussian decay envelope $\exp[-(t/T_2^*)^2]$ of free induction signals in bulk NV ensembles, which arises from quasi-static Gaussian magnetic field noise \cite{doherty2013,Pham2012,Barry2020RMP}. These studies show that the inhomogeneous broadening of the ODMR resonance and the corresponding $T_2^*$ distribution in bulk diamond are well described by a normal model, validating the assumption used in our simulation framework. 

\begin{figure}[bt]
    \centering
    \includegraphics[width=0.8\textwidth]{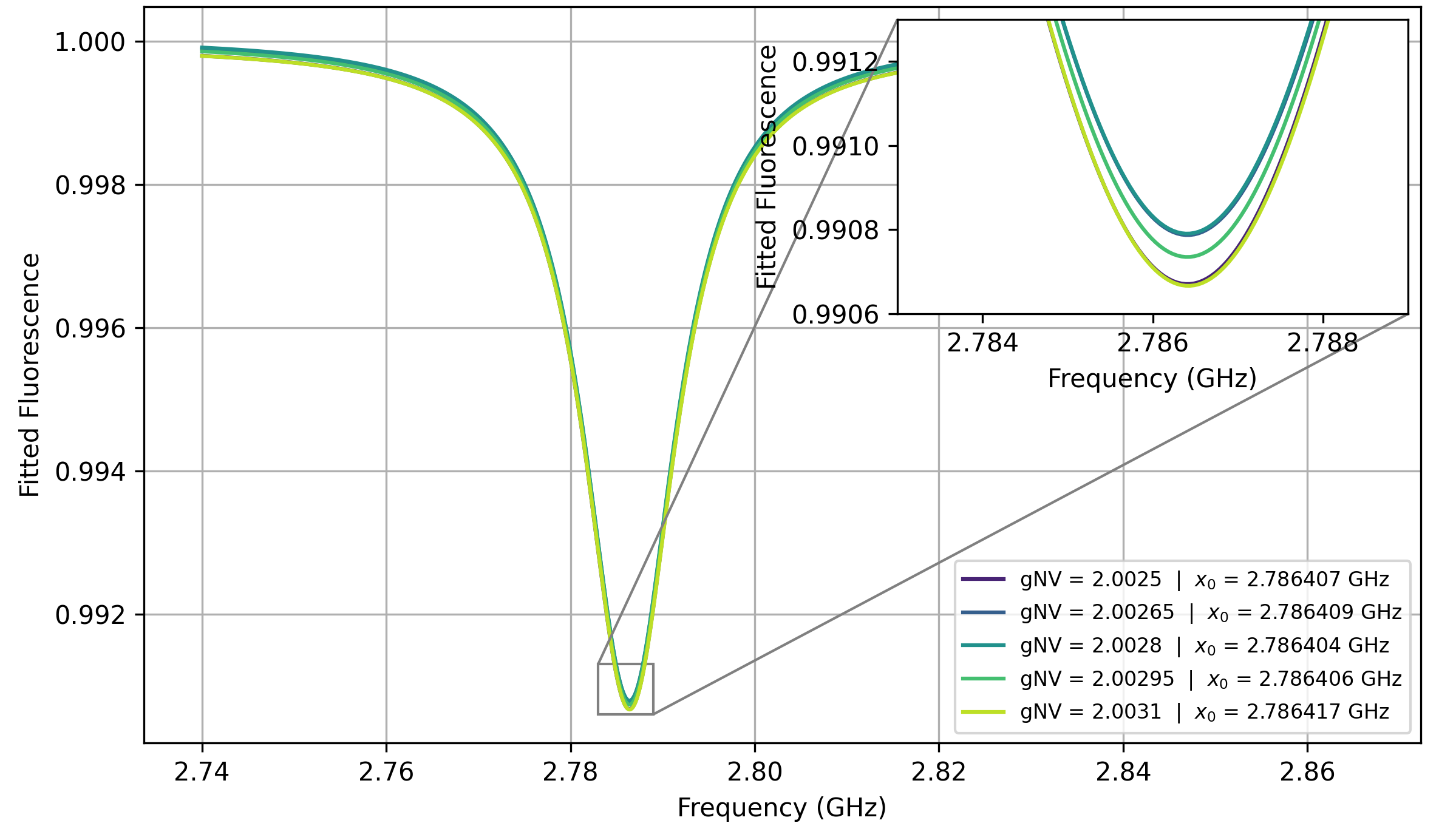}  
    \caption{ODMR spectrum at zero magnetic field for various values of the fluctuations in $g_{NV}$. The subplot show the shift in central peak due to different values of $g_{NV}$}
    \label{fig:g_NV}
\end{figure}
\subsection{Uncertainty of the gyromagnetic ratio.}
The effective \( g_{\mathrm{NV}} \) differs slightly from the free-electron gyromagnetic ratio due to the influence of the Coulomb potential from the nitrogen atom. Although the absolute value of \( g_{\mathrm{NV}} \) is well-characterized, there remains a small uncertainty in its precise value. To account for this, we model \( g_{\mathrm{NV}} \) as a random variable drawn from a normal distribution with a mean of 2.0028 and a standard deviation of 0.0003.

Calculation of the magnetic field using NV centers is linearly dependent on the gyromagnetic ratio of the NV center, defined as:
\begin{equation}
\gamma_{\mathrm{NV}} = \frac{g_{\mathrm{NV}} \mu_B}{h},
\end{equation}
where \( \mu_B \) is the Bohr magneton and \( h \) is Planck’s constant. The Bohr magneton is known to a very high precision, and uncertainty in its value is negligible, while the Planck constant is defined exactly \cite{RevModPhys.93.025010}. However, the effective \( g_{\mathrm{NV}} \)-factor for NV centers deviates from the free-electron \( g_e \)-factor due to the influence of the Nitrogen atom’s Coulomb potential \cite{Doherty_2011, PhysRevB.85.205203}. Additionally, the \( C_{3v} \) symmetry of the NV center allows for anisotropy in the Zeeman interaction, leading to different \( g \)-factors for magnetic fields parallel (\( g_{\parallel} \)) and perpendicular (\( g_{\perp} \)) to the NV axis. Because of these effects, \( g_{\mathrm{NV}} \) must be measured experimentally and carries a finite uncertainty. Assuming no anisotropy, the value \( g_{\mathrm{NV}} = 2.0028 \pm 0.0003 \) can be used, which contributes to the uncertainty in magnetic field estimation \cite{Lonard:2025ypd}. 

As the statistical distribution of the \( g_{\mathrm{NV}} \) value is not precisely known, we approximate it using a uniform distribution in our simulation, where a random value of \( g_{\mathrm{NV}} \) is selected in each iteration to incorporate the uncertainty in the gyromagnetic ratio. As we can in in \ref{fig:g_NV} the position of peaks are shifted depending on the value of $g_{NV}$ and thus the uncertainty in it's value can be a source of incorrect measurement.


\subsection{Laser intensity over an area}
\begin{figure}[tb!]
    \centering
    \includegraphics[width=0.5\textwidth]{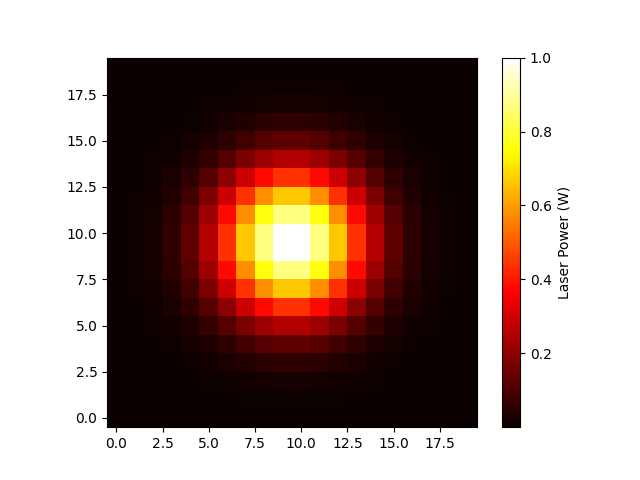}  
    \caption{Heatmap of Gaussian laser power distribution over the grid used for ODMR gnenration.}
    \label{fig:laser_grid}
\end{figure}

\begin{figure}[tb!]
    \centering
    \includegraphics[width=0.8\textwidth]{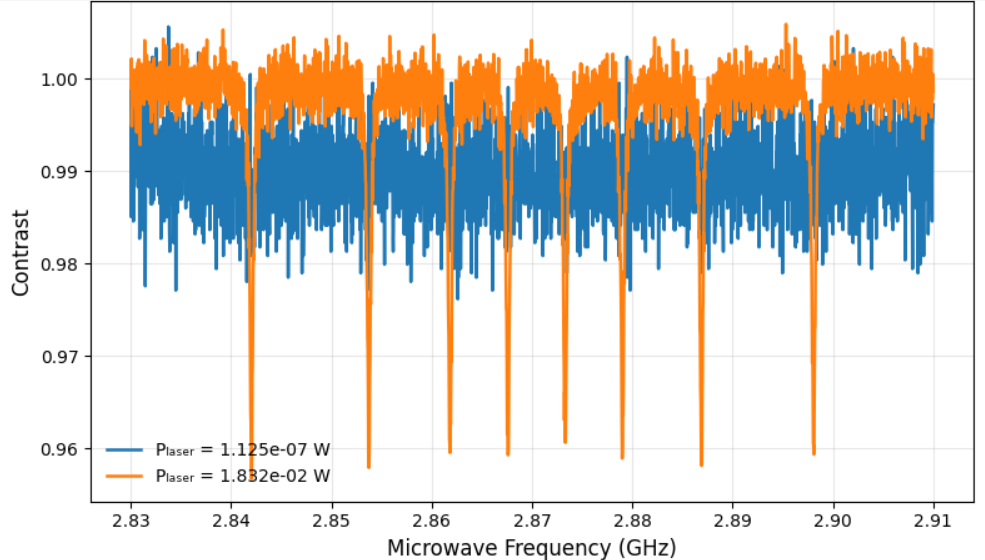}  
    \caption{Plot of ODMR at an near center point with relatively high laser power and at the edge with very small laser power. As we can see that the resolution gets destroyed as we approach the edges due the the laser power being very low there and the peak does not resolves in the ODMR.}
    \label{fig:nearfar_ODMR}
\end{figure}

\begin{figure}[tb!]
    \centering
    \includegraphics[width=0.8\textwidth]{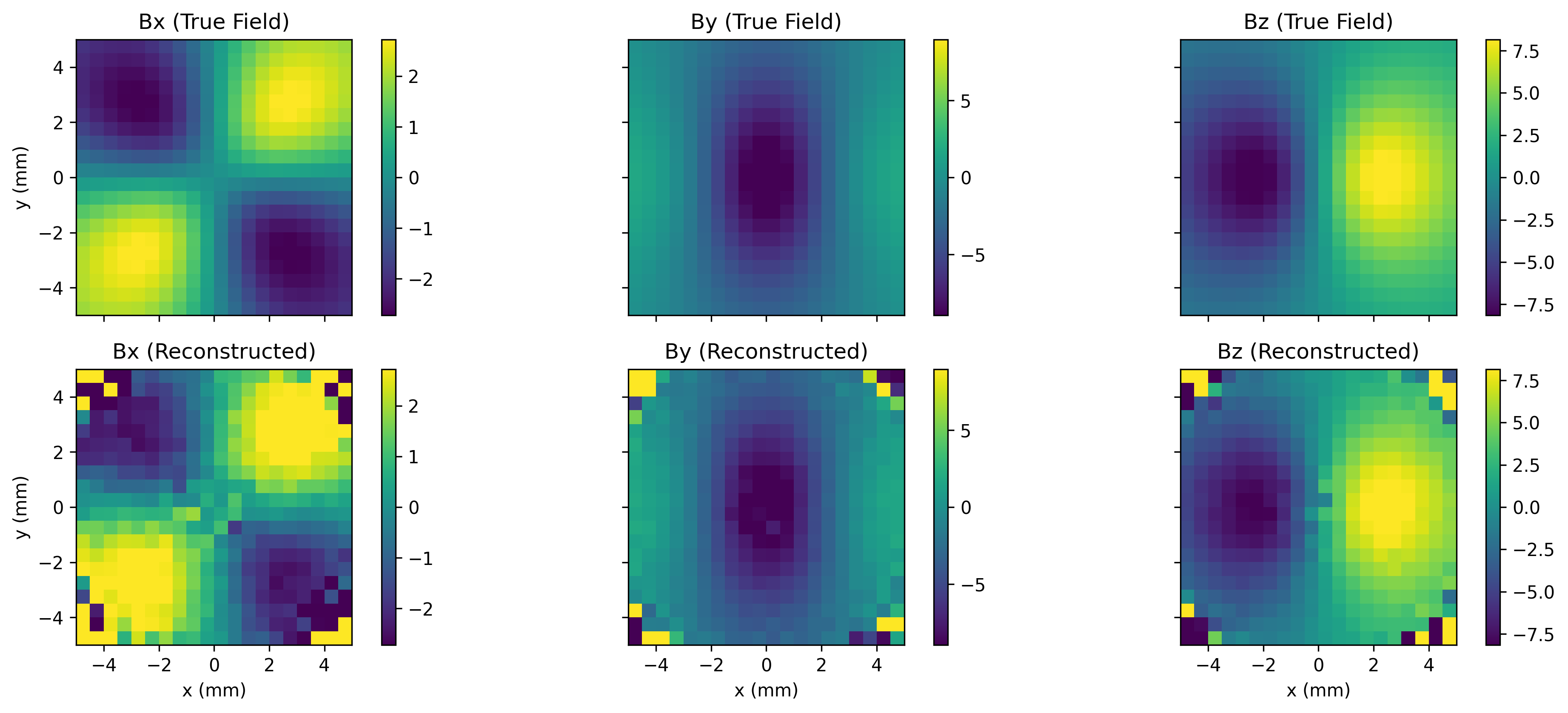}  
    \caption{Plot of magnetic field heatmaps, true and reconstructed, where the reconstructed fields are in the present of a Gaussian beam instead of uniform laser As we can see that the resolution gets destroyed as we aproach the edges due the the laser power being very low there.}
    \label{fig:nearfar_heat}
\end{figure}

While simulating the ODMR spectrum over a finite spatial region rather than at a single point, the laser intensity cannot be assumed to be uniform across the area. Instead, it exhibits a spatial variation that can be well approximated by a two-dimensional Gaussian profile~\cite{Siegman1986Lasers,Simpson2017Microscopy}. To account for this, we model the laser intensity \( I(x, y) \) at each point in the simulated 2D array as shown in \ref{fig:laser_grid} using the expression:

\[
I(x, y) = I_0 \exp\left(-\frac{2(x^2 + y^2)}{w_0^2}\right),
\]
where \( I_0 \) is the peak laser intensity at the center of the beam, and \( w_0 \) is the beam waist (the radius at which the intensity falls to \( 1/e^2 \) of \( I_0 \))~\cite{Siegman1986Lasers}. As we can see in Fig.\ref{fig:nearfar_ODMR} the ODMR spectrum is indistinguishable from noise near the end of the laser due to very small values of laser power and thus contrast values are too small to be observed as distinct from noise. The effect of this on magnetic field reconstruction can be seen in Fig.\ref{fig:nearfar_heat}

\subsection{Preferential orientation of NV defects}
\begin{figure}[tb!]
    \centering
    \includegraphics[width=1\textwidth, height = 0.5\textwidth]{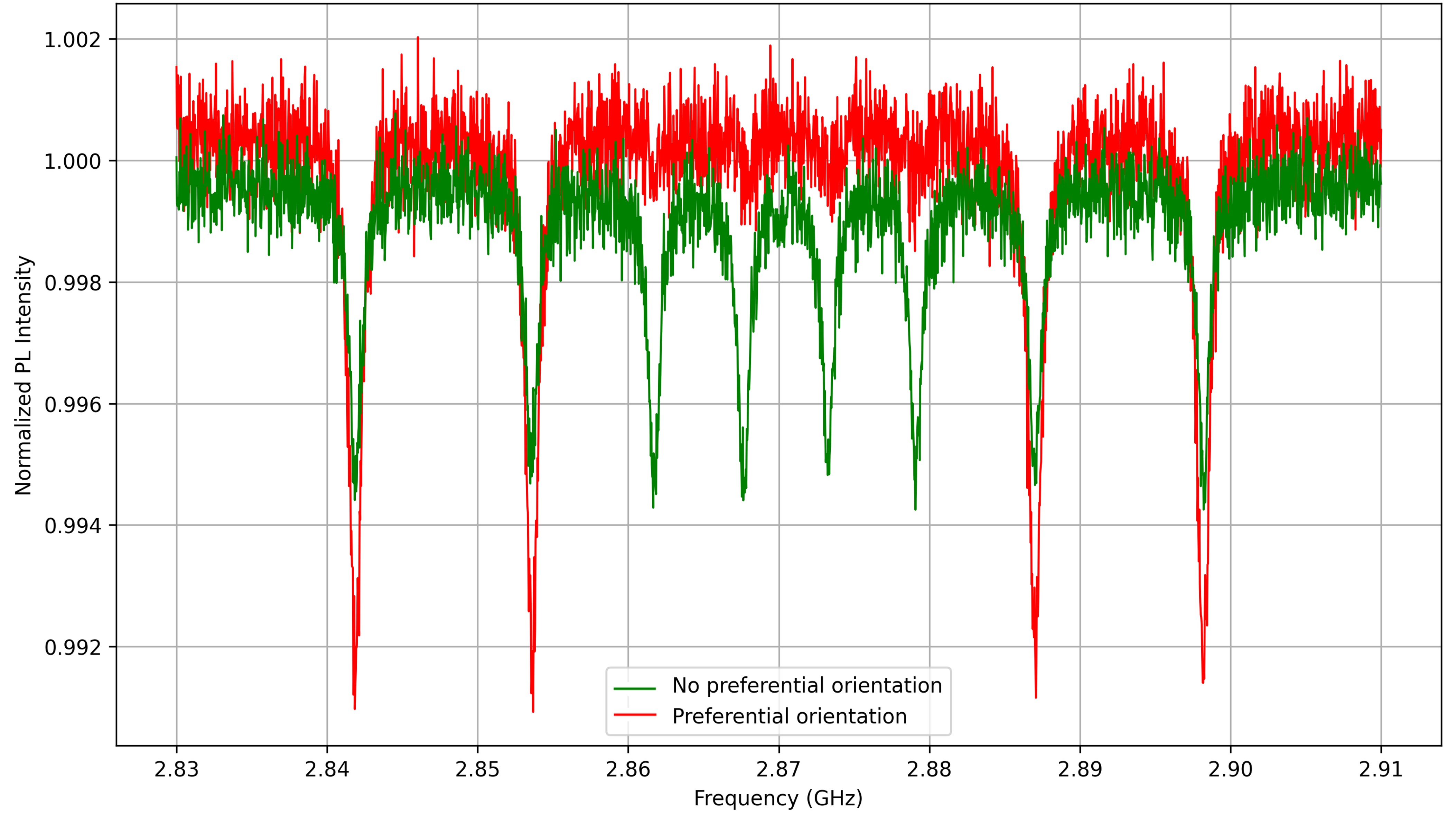} 
    \caption{Simulated ODMR spectra illustrating the effect of preferential orientation of NV centers. The green curve represents a case with no preferential orientation, where NV centers are evenly distributed along all four crystallographic axes. The red curve corresponds to a sample with \SI{94}{\percent} preferential orientation along two of the four crystallographic directions. leading to enhanced contrast and a reduction in spectral overlap. Such high degrees of orientation have been experimentally realized in (113)-oriented diamond samples~\cite{preferrential}.}
    \label{fig:preferential}
\end{figure}

NV centers can be classified based on the its orientation along one of the four crystallographic directions. In most samples the NV centers are present equally in all four direction \cite{preferrential}. It is shown that preferential orientation in only two direction can be realized on synthetically grown diamond \cite{twodirections}. Fig. \ref{fig:preferential} shows how the ODMR spectrum changes when there is preferential orientation of the NV centers. This results in increased contrast for the preferential  directions of NV centers which can be advantageous for implementations such as AC magnetic field sensitivity \cite{twodirections}.



\subsection{Magnetic Noise from Surface Impurities}
Surface impurities are known to introduce magnetic noise that affects the NV center spin dynamics. This noise arises primarily from spin flip and spin precession processes of paramagnetic impurity spins located in a thin surface layer \cite{Chrostoski2021}. The resulting temporal fluctuations in the local magnetic field lead to decoherence and spectral broadening in ODMR measurements.

To include the effect of surface magnetic noise in our simulation, we model the fluctuating surface magnetic field as a Gaussian-distributed random vector added to the external magnetic field experienced by each NV center \cite{}:
\[
\mathbf{B}_{\text{total}} = \mathbf{B}_{\text{ext}} + \mathbf{B}_{\text{surf}},
\]
where \( \mathbf{B}_{\text{surf}} \sim \mathcal{N}(0, \delta_B^2) \) represents the magnetic field fluctuations caused by surface impurities. The standard deviation \( \delta_B \) depends on the surface termination: for oxygen-terminated surfaces, the noise is relatively weak due to short impurity spin-lattice relaxation times (\(\tau_s \sim 7.5~\text{ps}\)), corresponding to \(\delta_B \sim 1\text{--}10~\text{nT}\). For hydrogen- and fluorine-terminated surfaces, the relaxation times are longer (\(\tau_s \sim 10\text{--}300~\text{\textmu s}\)), leading to stronger noise fields of \(\delta_B \sim 0.1\text{--}1~\text{\textmu T}\) \cite{Chrostoski2021}. 

Since our model employs an ensemble average for magnetic field calculations, the variance \( \delta_B^2 \) is scaled by the surface NV density, reflecting the uncorrelated summation of contributions from individual NVs. This stochastic treatment effectively captures the inhomogeneous broadening and decoherence due to surface spin noise, allowing us to simulate realistic ODMR spectra for NV ensembles near the diamond surface.

\section{Discussion}\label{sec:discussion}
\subsection{Reproduction of experimental results}
\begin{figure}[tb!]
    \centering
    \begin{subfigure}[b]{0.45\textwidth}
        \centering
        \includegraphics[width=\textwidth]{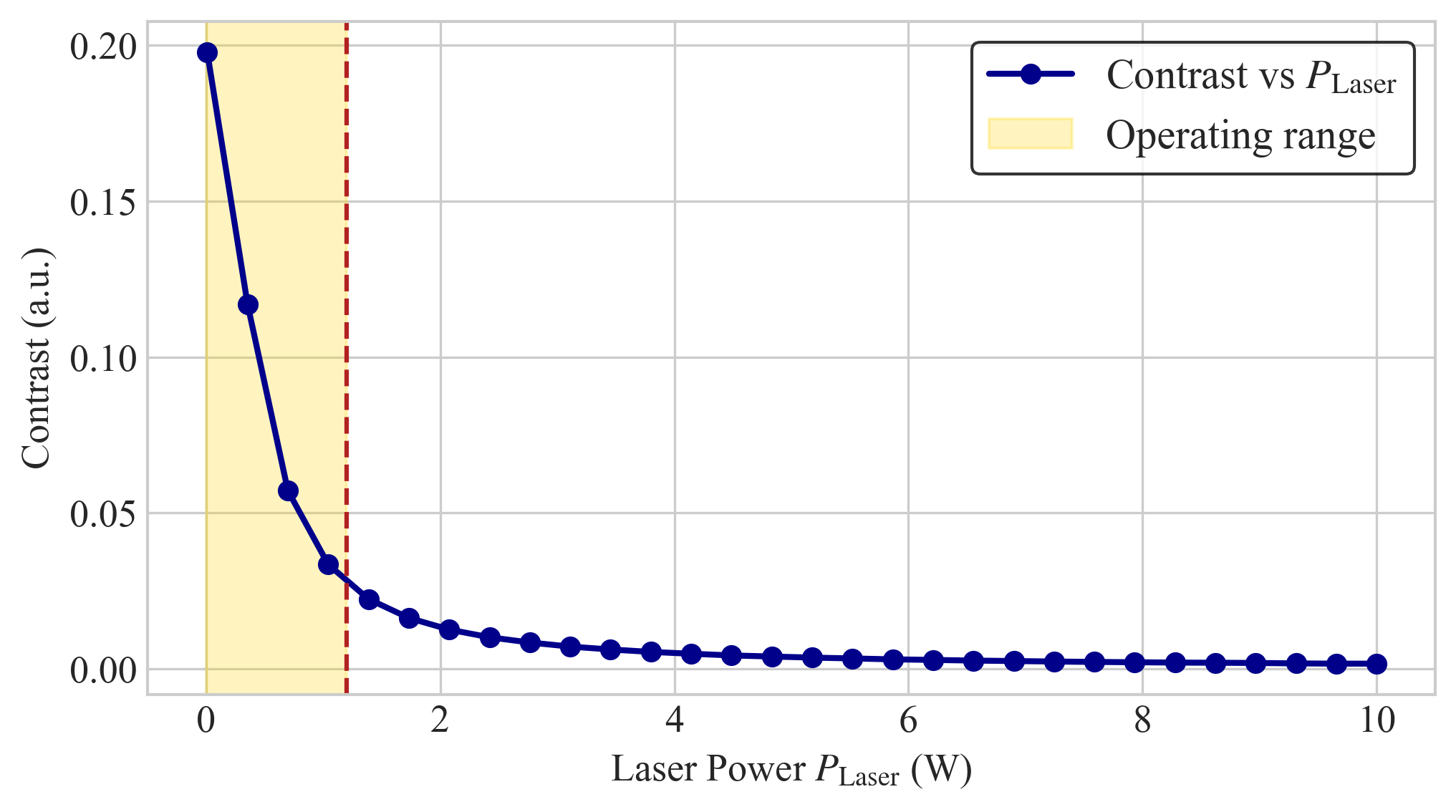}
        \caption*{\textbf{a}}
    \end{subfigure}
    \hfill
    \begin{subfigure}[b]{0.45\textwidth}
        \centering
        \includegraphics[width=\textwidth]{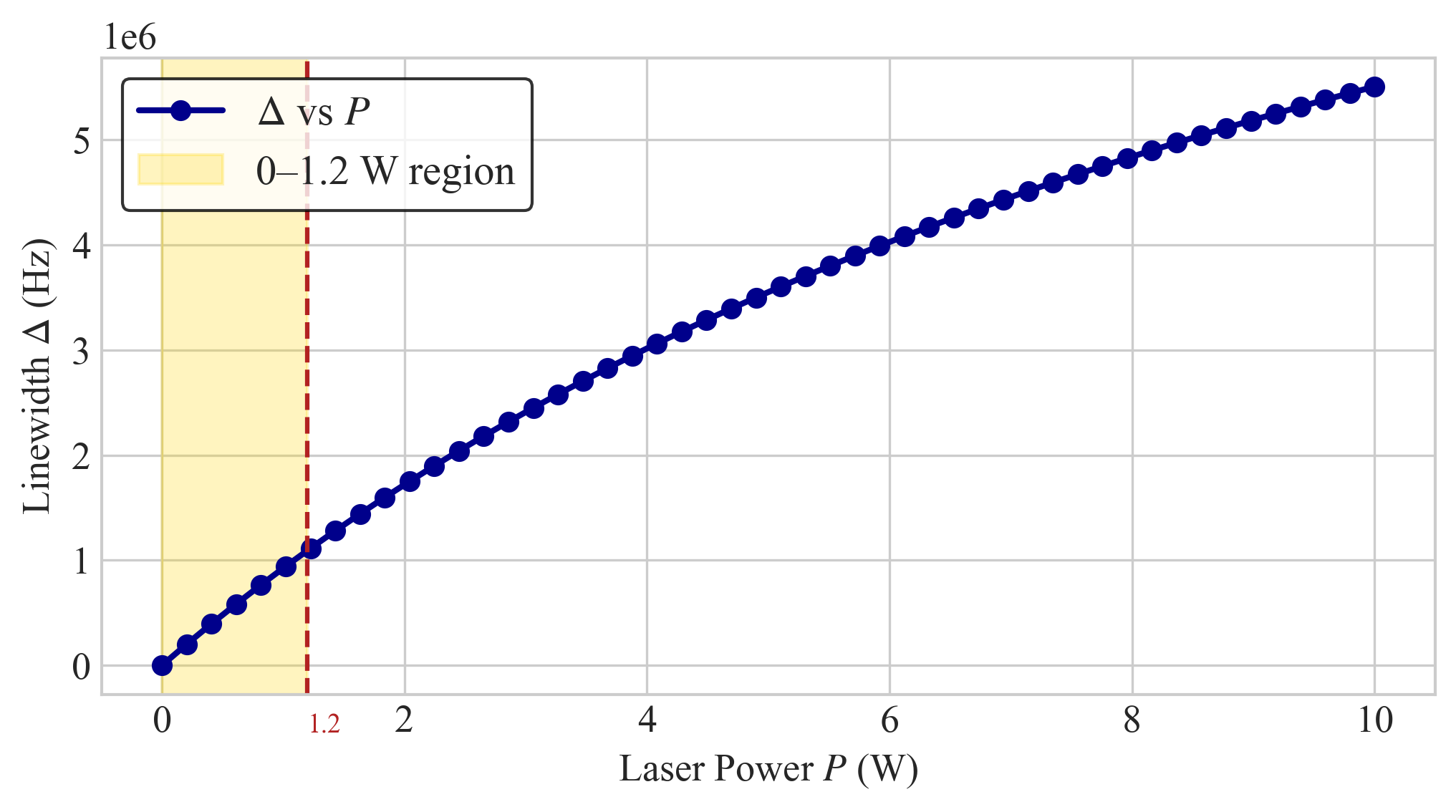}
        \caption*{\textbf{b}}
    \end{subfigure}

    \vskip 0.5em

    \begin{subfigure}[b]{0.45\textwidth}
        \centering
        \includegraphics[width=\textwidth]{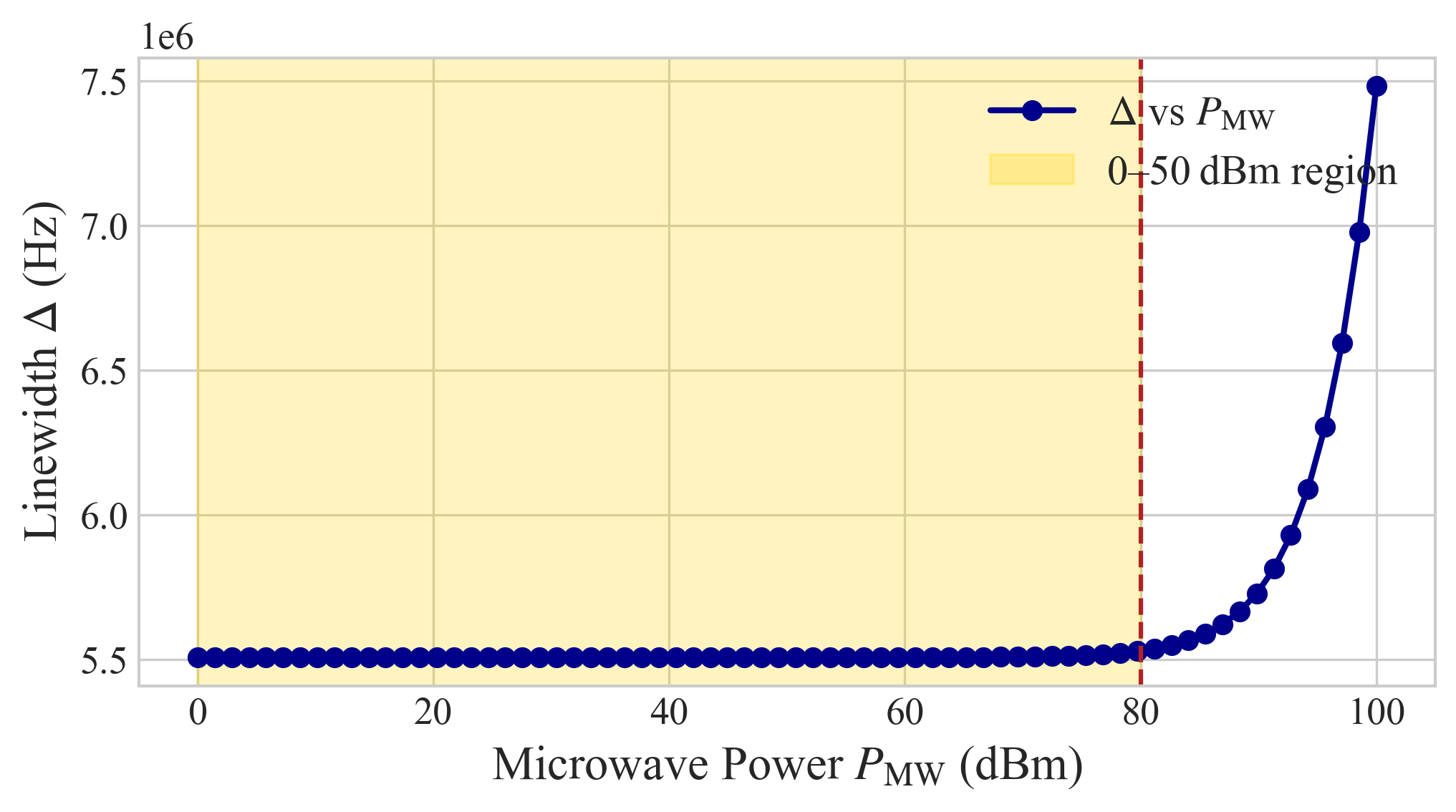}
        \caption*{\textbf{c}}
    \end{subfigure}
    \hfill
    \begin{subfigure}[b]{0.45\textwidth}
        \centering
        \includegraphics[width=\textwidth]{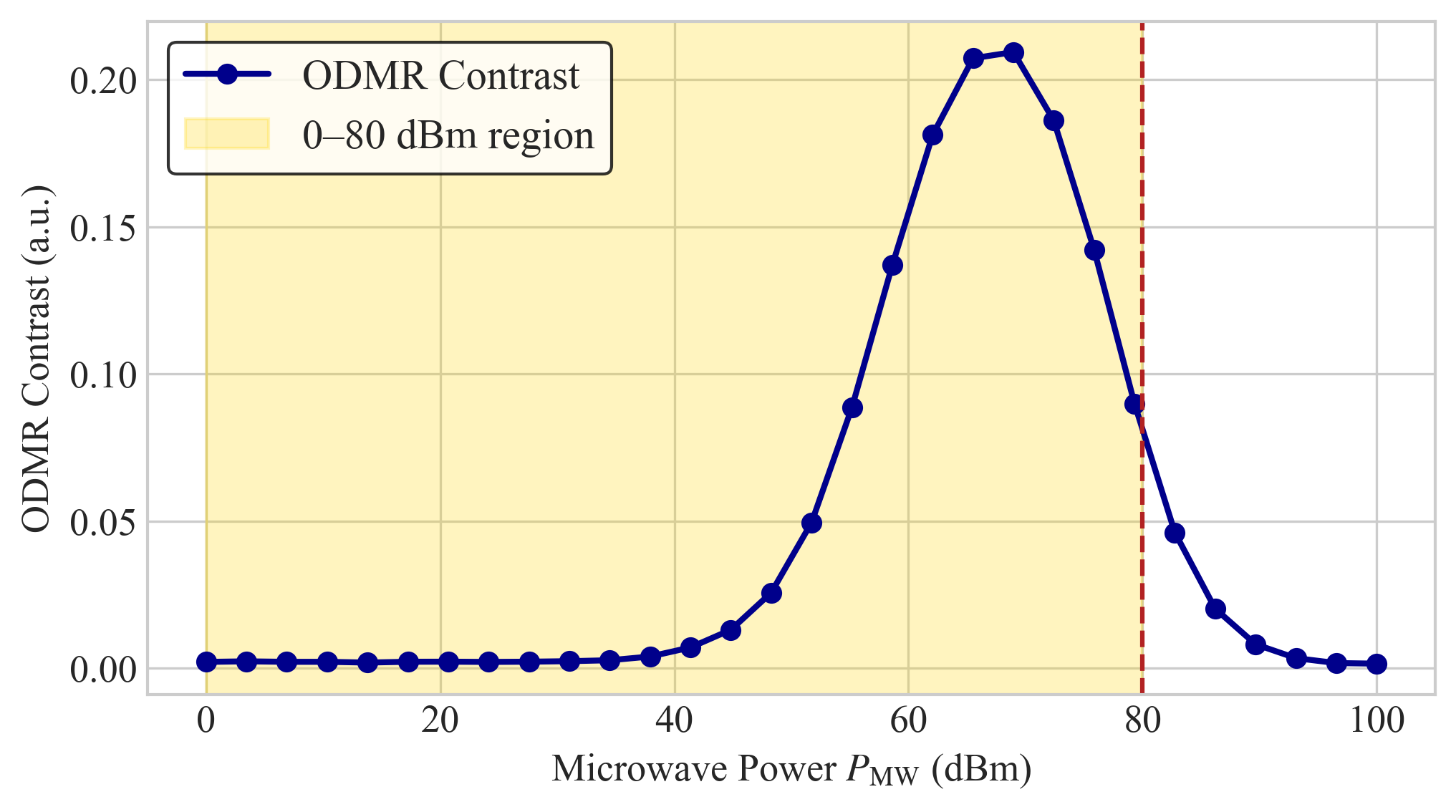}
        \caption*{\textbf{d}}
    \end{subfigure}

    \caption{(a)–(b) ODMR contrast and linewidth as functions of laser power; (c)–(d) ODMR contrast and linewidth as functions of microwave power. The yellow-shaded region indicates the operating range of power values reported in \cite{dreau2011avoiding}, from which certain parameters used in our simulations were adopted.}
    \label{fig:reproduction}
\end{figure}

To assess the validity of our simulation, we reproduced results with established experimental analogues. The behaviour of the ODMR contrast and linewidth under varying experimental parameters is illustrated in Fig. \ref{fig:reproduction}, and it demonstrates qualitative agreement with the trends reported in \cite{dreau2011avoiding}, thereby supporting our model.
As shown in Fig.~\ref{fig:reproduction} (a), though the ODMR contrast initially increases with laser power due to more efficient optical pumping into the $m_s = 0$ state \cite{doherty_nv_review}, which enhances the spin polarization responsible for the resonance signal. The contrast stabilizes and then decreases with further increase in due to saturation
and increased background fluorescence, reducing the visibility of the ODMR dip \cite{doherty_nv_review,Rondin2014}. Simultaneously, Fig.~\ref{fig:reproduction} (b) and (c) shows that the linewidth of the resonant dips increases monotonically with laser and microwave power, a behaviour primarily attributed to optical and microwave-induced power broadening respectively along with additional dephasing \cite{Tetienne_2012}.

A similar trend is observed when the microwave power is varied. As shown in Fig.~\ref{fig:reproduction} (d), increasing the microwave power initially enhances the ODMR contrast, since the driving field becomes more effective in inducing transitions between the spin sublevels. However, beyond a certain threshold, further increases in microwave power lead to saturation of the spin transition,  and any further increase degrades the contrast due to power broadening. \cite{dreau2011avoiding}.


\subsection{Parameter Optimization}

One of the key applications of the developed simulation framework is to determine the optimal operating conditions for the laser and microwave powers that yield the most clearly resolvable ODMR signal. For reliable measurements, it is desirable to operate in a regime where the contrast is as high as possible for a narrow linewidth, allowing the resonance dip to be distinctly separated from background noise and clearly resolvable \cite{Barry2020RMP}.

However, as observed from the simulation results, these two parameters—contrast and linewidth—are not independent \cite{dreau2011avoiding}. Their relationship is inherently complex, and varying one does not necessarily lead to a predictable change in the other. For instance, increasing the microwave power may improve the contrast up to a certain point, but it can also cause power broadening, which increases the linewidth. This interdependence highlights the importance of identifying an optimal balance between the two.

To systematically quantify this trade-off, we define a figure of merit (FOM) as
\begin{equation}
\text{FOM} = \frac{\text{Contrast}}{\text{Linewidth}},
\end{equation}
which effectively captures the balance between signal distinguishability and spectral resolution. Maximizing this ratio provides an objective way to identify the conditions under which a resonant frequency is best resolved \cite{Broadway2018PRApplied}.

\begin{figure}[tb!]
    \centering
    \includegraphics[width=0.8\textwidth]{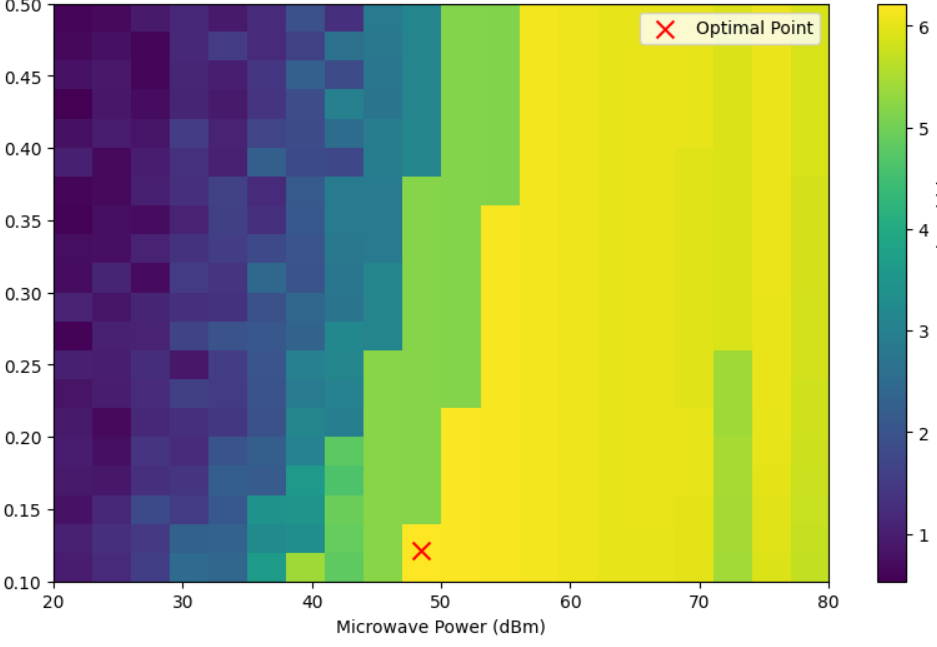}  
    \caption{Heatmap of the figure of merit (FOM = Contrast/Linewidth) as a function of laser and microwave powers. The color scale represents the relative magnitude of the FOM, with brighter regions indicating the optimal operating conditions, demonstrating a systematic way to tackle the trade-off between ODMR contrast and linewidth.}
    \label{fig:parameter}
\end{figure}

Using the developed simulation, we can evaluated the FOM over a broad range of experimentally accessible laser and microwave powers. This analysis enables efficient parameter estimation by revealing the operating region that offers the best compromise between contrast and linewidth for a specific experimental setup encoded through other hyperparameters. The resulting heatmap of the figure of merit across the parameter space is shown in Fig.~\ref{fig:parameter} for an instance of simulation parameters, which clearly indicates the optimal zone for achieving maximum ODMR resolvability.

\subsection{Denoising}
From a typical noisy ODMR, the resonant frequencies and in turn, magnetic field values can be obtained by solving a multi-parameter estimation problem \cite{phasebased}. But this is majorly limited, for instance as given in Section \ref{sec:odmr-method}, by the initial guess-values for the non-linear fitting of Lorenztians for certain levels of SNR. Moreover, the fitting process becomes unreliable beyond a certain SNR value, necessitating identification and removal of noise from ODMR spectra. 

Here, we utilized a few classical techniques of denoising 1D signals and brought out the relevance of them by comparing the accuracy enhancements. We employed three simple techniques; Gaussian denoising (GD), Bilateral filtering (BF) \cite{edge} and Wavelet denoising (WD) \cite{pywt} independently and sequentially for each pixel's ODMR. Then pertaining to the 3D data-structure of ODMR for all pixels in a wide-field setup, we employed off-the-shelf denoiser, Block-matching and 4D filtering (BM4D), to denoise the ODMR data-cube in one-shot. BM4D is an extension of a more popular BM3D denoising algorithm, used for 2D multi-channel data \cite{bm4d}. 

\begin{figure}[ht]
    \centering 
    \begin{subfigure}[b]{0.67\linewidth}
        \label{odmrsecden}
\includegraphics[width=\textwidth]{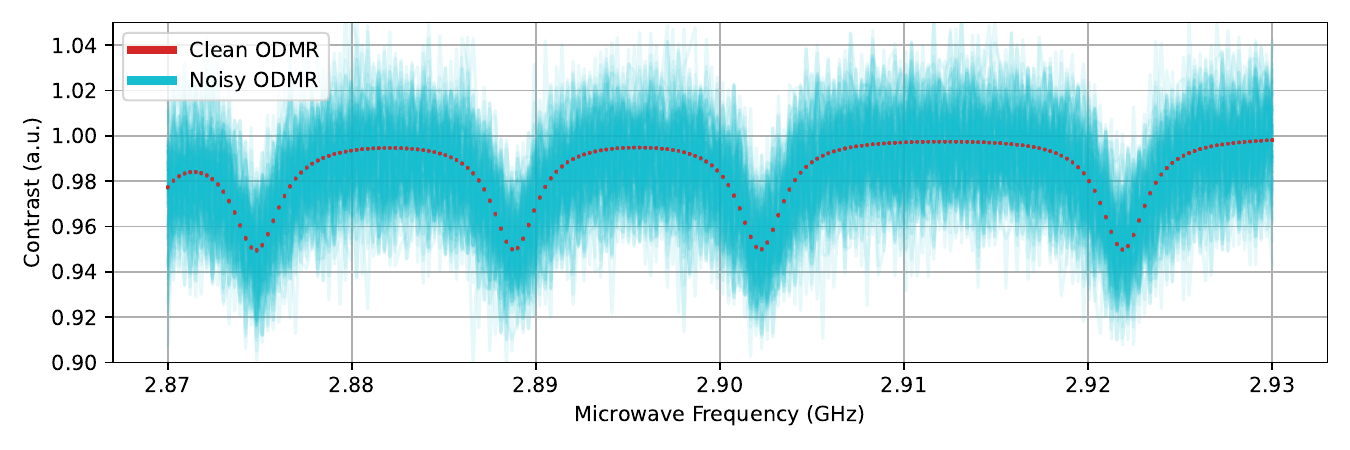}
    \end{subfigure}
    \hfill
    \begin{subfigure}[b]{0.3\linewidth}
        \label{snrsecden}
\includegraphics[width=\textwidth]{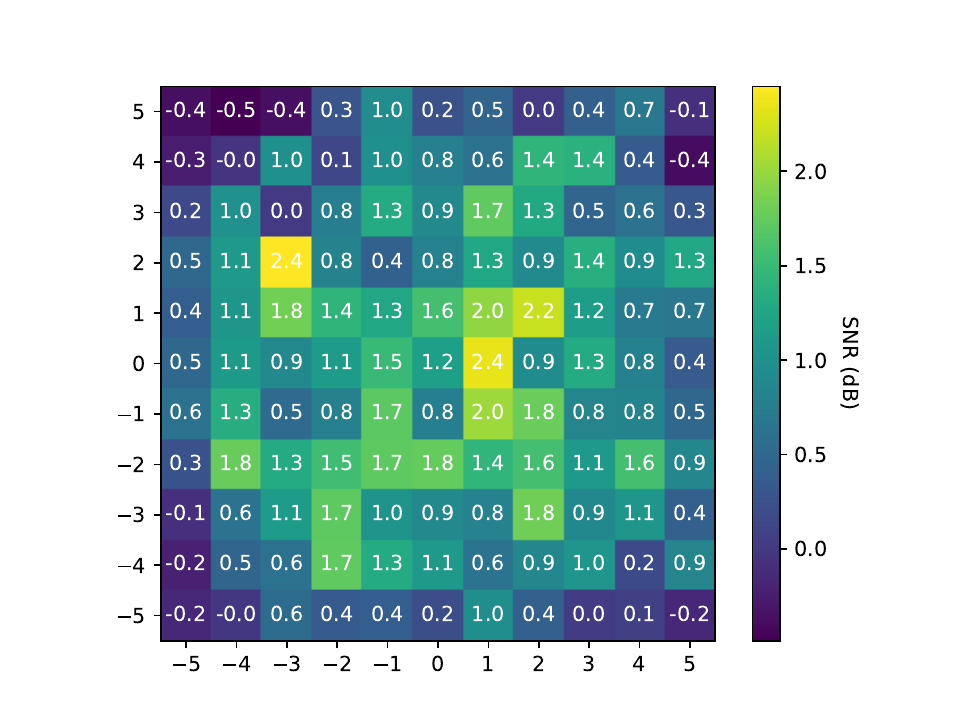}
    \end{subfigure}    
    \caption{\protect \subref{odmrsecden} A set of typical ODMRs with realistic inhomogeneities and noise contribution for a perfectly homogeneous magnetic field. We assume a gaussian LASER beam incident on the diamond surface with beam waist of 11 \textmu m and a loop MW antenna producing homogeneous MW field along z-axis, across this small area with a frequency interval of 300 kHz. Both fields are modeled to fluctuate with a standard deviation of $6.5\%$ relative to the mean power. \protect \subref{snrsecden} SNR calculated for each ODMR instance is presented. Average SNR of the ODMR response is 0.85 dB with a standard deviation of 0.62 dB. We can see a strong correlation between SNR and LASER power distribution. }
    \label{denoisedata}
\end{figure}

A bias-magnetic field $\bm{B}_{bias}$ was chosen so that all eight resonant frequencies are fully resolved, as described in Fig. \ref{denoisedata}. A small sample magnetic field $\bm{B} = 50 \text{\textmu T} \ \hat{z}$ was assumed, shifting resonant frequencies marginally for input $\bm{B}_{with\ sample} = \bm{B}_{bias}+\bm{B}$ (Fig. \ref{denoisedata}). Note that the shift is well-resolvable in these ODMR instances with frequency step of 300 kHz, corresponding to $\approx 10.7 \text{\textmu T}$. If we directly do a non-linear fitting using built-in scipy function based on a Trust-region reflective algorithm \cite{trf}, an average NRMSE of $(565.8\pm 0.05) \times 10^{-3}$ is achieved in magnetic field estimation $\tilde{\bm{B}}$. Note that the performance depends on the quality of initial guesses and hence they are kept same for all the experiments in this section, chosen on the basis of visual inspection.

\subsubsection{Gaussian Filter}
We start with a weighted linear smoothening process, involving a convolution with a Guassian kernel \cite{edge} and our noisy ODMR signal. The only hyperparameter for gaussian kernel is its standard deviation $\sigma$ and we chose the following values for the Table \ref{table:denoise} $\mu = \sigma \in \{ 0.5, 1.0,1.5 , 2.0,2.5,3.0,3.5,4.0 \}$. From Table \ref{table:denoise}, one can notice an optimized $\sigma$ for the kernel that resulted into the highest SNR. We also note the NRMSE of the relevant quantities (resonant frequencies) obtained from the same fitting procedure on the denoised dataset and map them to respective $\sigma$.  

\subsubsection{Wavelet Denoising}
An efficient discrete wavelet transform (DWT) brings the data to transform domain where the actual signal can have a sparse representation and a subsequent filtering can be done based on a threshold. It was found out that Daubechies and Biorthogonal wavelets perform robustly for denoising purposes on ODMR, db11 being the best. Python based module Pywavelet \cite{pywt} for the efficient DWT and thresholding in garrotte mode and inversion was utilized. We varied the thresholding value $\mu \in \{ 0.001,0.003, 0.009, 0.026, 0.077, 0.228, 0.675, 2.000 \}$ in Table \ref{table:denoise} and note the performance. Note that the SNR saturates after a $\mu = 0.077$ for the chosen mode of DWT and couldn't outperform GF. One can also use continuous Lorenztian or Gaussian type wavelets from the same module for a complete wavelet analysis which are more appropriate for our case. This can perform thorough noise removal for small dataset but is impractical for handling real datasets. 

\subsubsection{Bilateral Filter}
In this section, a 1D version of bilateral filtering that is regularly used in image denoising with good edge preservation \cite{edge} is employed for the task. It can smoothen the signal not only based on the nearby points like GF but also takes in account of the ODMR contrast values, bringing anisotropy. Since our ODMR signal can also be divided into different regions based on the contrast values, BF can help preserving the dips of the Lorentzians. We assumed gaussian functions for both range kernel and spatial kernel with $\sigma_r = 1. $ fixed and $\mu = \sigma_s$ taking values $\sigma_s \in \{0.5,1.0,1.5 , 2.0,2.5,3.0,3.5,4.0\}$, being their standard deviations respectively. We kept the optimized window length $\Omega=20$, same for all samples. For larger window lengths $\Omega>25$, the denoising behavior convergences to that of GF, expectedly. Table \ref{table:denoise} shows an improvement in SNR with respect to GF but at the cost of large computational time. 

\begin{table}[tb!]
\centering
\caption{A summary of performance of various smoothening techniques with varying parameters is presented here. We track SNR of denoised ODMR signal and corresponding NRMSE of resultant resonant frequencies obtained after non-linear fitting with the same hyper-parameters. We noted a robust and efficient performance from GF in this scenario among all three techniques.}
\label{table:denoise}
\begin{tabular}{@{}clccccllll@{}}
\toprule
\begin{tabular}[c]{@{}c@{}}Filtering\\  Technique\end{tabular} & Metrics  & \multicolumn{8}{c}{Denoising Hyperparameter ($\mu$)}  \\ \midrule
\multirow{2}{*}{GD}                                            & SNR (dB) & 2.55 & 5.92 & 7.44 & 8.35 & 8.86 & 9.08 & 9.09 & 8.97 \\
                                                               & NRMSE    & 0.62 & 0.52 & 0.51 & 0.51 & 0.54 & 0.55 & 0.57 & 0.58 \\ \midrule
\multirow{2}{*}{WD}                                            & SNR (dB) & 0.70 & 0.84 & 1.79 & 6.03 & 8.34 & 8.34 & 8.34 & 8.34 \\
                                                               & NRMSE    & 0.58 & 0.62 & 0.58 & 0.53 & 0.57 & 0.57 & 0.57 & 0.57 \\ \midrule
\multirow{2}{*}{BF}                                            & SNR (dB) & 2.55 & 5.92 & 7.45 & 8.36 & 8.87 & 9.09 & 9.10 & 8.99 \\
                                                               & NRMSE    & 0.62 & 0.54 & 0.52 & 0.50 & 0.54 & 0.55 & 0.57 & 0.58 \\ \bottomrule
\end{tabular}

\end{table}


\subsubsection{BM4D}
\begin{figure}[tb!]
    \centering 
    \begin{subfigure}[b]{0.6\linewidth}
    \includegraphics[width=\textwidth]{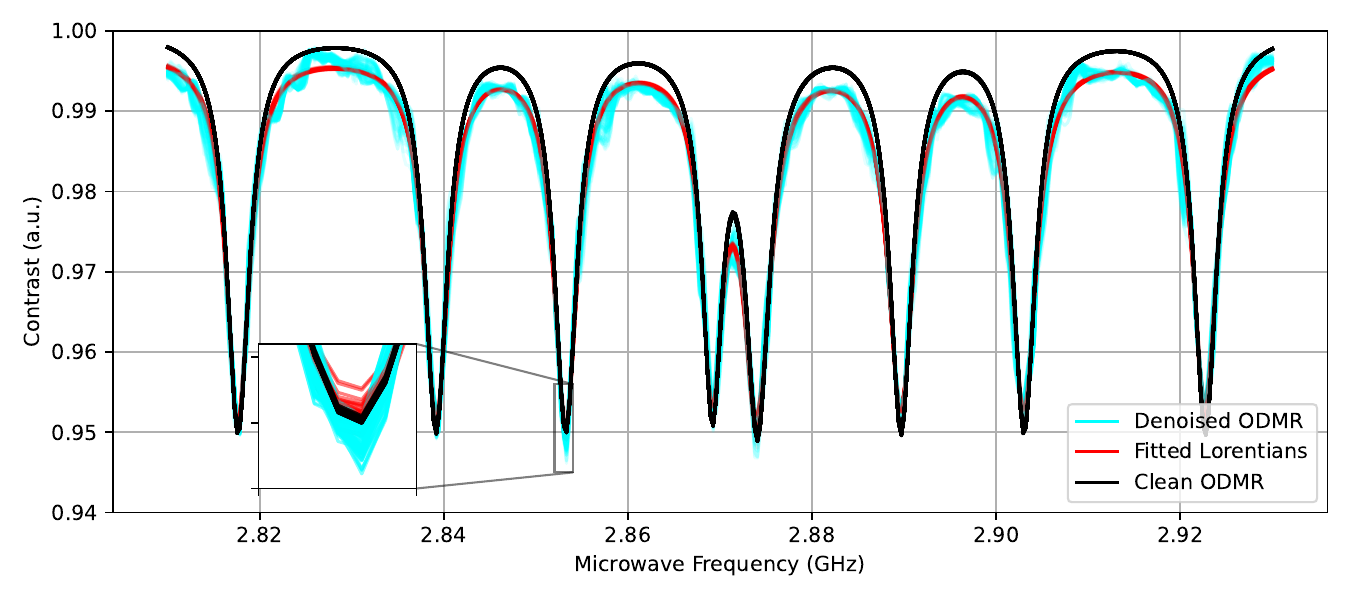}
    \label{odmrsecdenoise}
    \end{subfigure}
    \hfill
    \begin{subfigure}[b]{0.36\linewidth}
    \includegraphics[width=\textwidth]{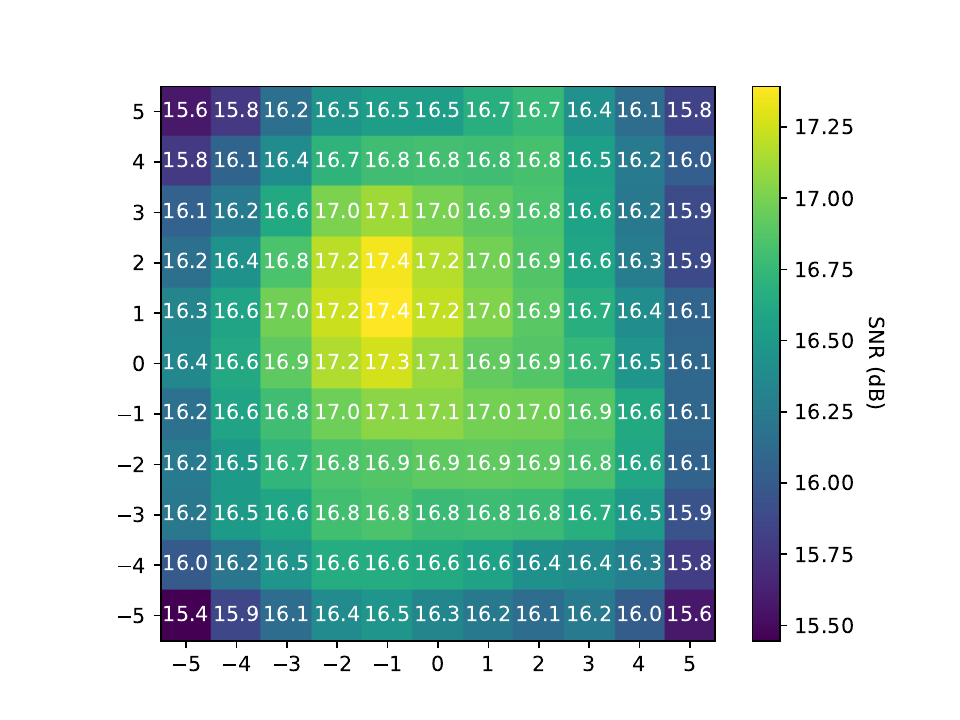}
    \label{snrsecdenoise}
    \end{subfigure}    
    \caption{\subref{odmrsecdenoise} The comparison of noisy ODMR instance with their BM4D-denoised counterparts. We observed an average SNR for denoised signals to be 16.66 dB with 0.079 NRMSE providing a dramatic performance enhancement compared to 1D filtering techniques. \subref{snrsecdenoise} SNR calculated for each denoised ODMR instance is presented. We can again see its strong correlation with LASER power distribution.}
    \label{bm4d}
\end{figure}

Changing the perspective from ODMR as multiple 1D signals for multiple pixels of camera to ODMR as a single data cube, we can utilize the correlations between adjacent ODMRs. It can be intuitively motivated by considering the continuous nature of magnetic field vectors, making resonant frequencies of nearby pixels closer to each other. This further helps due to the fact that there are finite noise correlations across all three axes. We used a popular denoiser for 3D data structure BM4D with a single parameter of noise variance $\mu= 0.02$ for this particular dataset. We see a dramatic improvement in both SNR and NRMSE as can be seen from Fig. \ref{bm4d}. It is to be noted that we can attribute this improvement to the almost perfect correlations (Fig. \ref{bm4d}) between ODMRs belonging to different camera pixels due to both homogeneous bias magnetic field and sample magnetic field. When added a little inhomogeneity in the sample magnetic field, SNR of denoised ODMRs showed an average decrement of 0.5 dB. This needs to be further studied and would be the scope of our future work.

We can further bring out the relevance of this perspective by comparing BM4D's operation with BM3D operation on each fluorescence intensity image. Applying BM3D on each image individually and stacking them again as datacube gives a similar performance of 15.69 dB SNR and 0.056 NRMSE but in $\sim 80 \times$ more time on the same machine. Hence, processing ODMR as data cubes seems to be a promising frontier and further motivates us to employ data driven approaches trained on the simulated results from our framework.

\section{Conclusion}\label{sec:conclusion}
We have developed a comprehensive and physically grounded simulation framework for continuous-wave ODMR in NV center ensembles, designed to faithfully capture the behavior of realistic experimental systems. By combining a seven-level description of NV spin dynamics with analytically and numerically modeled noise sources, the framework functions as a digital twin that bridges the gap between idealized theory and practical implementation and can vastly speed-up the deployement of these quantum sensors for various of applications ranging from healthcare to electronics.

The model incorporates a broad spectrum of experimentally relevant imperfections, including laser and microwave power fluctuations, microwave phase noise, spin dephasing, temperature-induced shifts in the zero-field splitting, surface-related magnetic perturbations, uncertainty in the NV gyromagnetic ratio, and photon shot noise. Power broadening and contrast reduction are treated self-consistently through linewidth calculations, while spatial inhomogeneity across the sensing region is captured via a Gaussian laser intensity profile. Crucially, all noise mechanisms are parameterized using quantities directly accessible in laboratory settings, allowing for a transparent mapping between simulation inputs and experimental conditions.

We further illustrate the practical utility of the framework by optimizing the contrast-to-linewidth ratio, a key figure of merit that governs magnetic-field sensitivity. These results highlight the digital twin's capability to guide experimental parameter selection and diagnose performance limitations arising from competing noise processes.

Beyond its immediate application to CW ODMR magnetometry, the framework is readily extensible. Future developments may include explicit treatment of electric-field and strain effects, incorporation of hyperfine interactions, aberrations arising due to the optical assembly, adaptation to pulsed ODMR protocols and time-domain sensing schemes, and inclusion of add-on electronics for active closed-loop control mechanisms. This can further help this toolkit to act as a testbed for applying various machine learning and signal processing techniques. As such, the present work provides a robust foundation for simulation-driven design and optimization of NV-based quantum sensors, supporting their continued evolution toward reliable, high-performance, and field-deployable technologies.



\bibliography{citations}
\end{document}